\documentclass[twocolumn]{aastex61}
\usepackage{longtable}
\usepackage{graphicx}
\usepackage{amsmath,amssymb}
\usepackage{booktabs}
\usepackage{color}
\usepackage{units}
\usepackage{epstopdf}
\usepackage[caption=false]{subfig}
\usepackage{float}
\usepackage{multirow}

\newcommand{\beq}{\begin{equation}}
\newcommand{\eeq}{\end{equation}}
\newcommand{\bdm}{\begin{displaymath}}
\newcommand{\edm}{\end{displaymath}}

\newcommand{\msun}{M_{\odot}}
\definecolor{Gray}{gray}{0.9}
\definecolor{orange}{rgb}{0.9,0.5,0}
\definecolor{dkGreen}{rgb}{0.09, 0.45, 0.27}
\definecolor{purple}{rgb}{0.44, 0.0, 1.0}

\newcommand{\LVC}{LIGO/Virgo}

\begin{document}
\title{Low-Latency Gravitational Wave Alerts for Multi-Messenger Astronomy During the Second Advanced LIGO and Virgo Observing Run}


\author{B.~P.~Abbott}
\affiliation{LIGO, California Institute of Technology, Pasadena, CA 91125, USA}
\author{R.~Abbott}
\affiliation{LIGO, California Institute of Technology, Pasadena, CA 91125, USA}
\author{T.~D.~Abbott}
\affiliation{Louisiana State University, Baton Rouge, LA 70803, USA}
\author{S.~Abraham}
\affiliation{Inter-University Centre for Astronomy and Astrophysics, Pune 411007, India}
\author{F.~Acernese}
\affiliation{Universit\`a di Salerno, Fisciano, I-84084 Salerno, Italy}
\affiliation{INFN, Sezione di Napoli, Complesso Universitario di Monte S.Angelo, I-80126 Napoli, Italy}
\author{K.~Ackley}
\affiliation{OzGrav, School of Physics \& Astronomy, Monash University, Clayton 3800, Victoria, Australia}
\author{C.~Adams}
\affiliation{LIGO Livingston Observatory, Livingston, LA 70754, USA}
\author{R.~X.~Adhikari}
\affiliation{LIGO, California Institute of Technology, Pasadena, CA 91125, USA}
\author{V.~B.~Adya}
\affiliation{Max Planck Institute for Gravitational Physics (Albert Einstein Institute), D-30167 Hannover, Germany}
\affiliation{Leibniz Universit\"at Hannover, D-30167 Hannover, Germany}
\author{C.~Affeldt}
\affiliation{Max Planck Institute for Gravitational Physics (Albert Einstein Institute), D-30167 Hannover, Germany}
\affiliation{Leibniz Universit\"at Hannover, D-30167 Hannover, Germany}
% ...

\collaboration{The LIGO Scientific Collaboration and the Virgo Collaboration}




\date{\today}

\begin{abstract}
Advanced LIGO's second observing run (O2), conducted from November 30, 2016 to August 25, 2017, combined with Advanced Virgo's first observations in August 2017 witnessed the birth of gravitational-wave multi-messenger astronomy. The first ever gravitational-wave detection from the coalescence of two neutron stars, GW170817, and its gamma-ray counterpart, GRB 170817A, led to an electromagnetic follow-up of the event at an unprecedented scale. Several teams from across the world searched for EM/neutrino counterparts to GW170817, paving the way for the discovery of optical, X-ray, and radio counterparts. In this article, we describe the online identification of gravitational-wave transients and the distribution of gravitational-wave alerts by the LIGO and Virgo collaborations during O2. We also describe the gravitational-wave observables which were sent in the alerts to enable searches for their counterparts. Finally, we give an overview of the online candidate alerts shared with observing partners during O2. Alerts were issued for 14 candidates, six of which have been confirmed as gravitational-wave events associated with the merger of black holes or neutron stars. Eight of the 14 alerts were issued less than an hour after data acquisition.
\end{abstract}

\section{Introduction}
\label{sec:intro}
Gravitational-wave (GW) multi-messenger astronomy provides a unique view of the cosmos. In this paper, we explain the procedures used during the second observing run of the advanced ground-based gravitational-wave-detector network to issue alerts for multi-messenger follow-up. We also include a summary of all alerts issued to observing partners and an update on the status of candidate events.

The Advanced Laser Interferometer Gravitational-wave Observatory (LIGO) detectors (\citealt{2015CQGra..32g4001L}) are installed in the US at Hanford, WA and Livingston, LA while the Advanced Virgo detector (\citealt{2015CQGra..32b4001A}) is located in Cascina, Italy near Pisa. The detectors use a modified Michelson laser interferometer design to measure GW strain. A passing GW causes a differential length change in the detector arms, producing interference of the laser beams at the beam splitter, and giving an optical readout proportional to the GW strain.

In September 2015 the two Advanced LIGO detectors began their first observing run (O1), lasting four months. The first direct detection of gravitational waves, GW150914, from the coalescence of binary black holes (BBH;  \citealt{2016PhRvL.116f1102A}) marked the beginning of gravitational-wave (GW) astronomy. Two additional BBH merger signals, GW151012\footnote{The candidate LVT151012 has been confirmed as a gravitational-wave event, now called, GW151012 \citep{CBCcatalog}.} \citep{2016PhRvX...6d1015A} and GW151226 \citep{2016PhRvL.116x1103A}, were identified before the end of O1. Following hardware and software upgrades, the second Advanced LIGO observing run (O2) began on 30 November 2016. Advanced Virgo joined  the network in August 2017 for the last month of data acquisition.

A number of additional BBH coalescences were detected in O2 (see \citealt{2017ApJ...851L..35A,2017PhRvL.119n1101A,2017PhRvL.118v1101A,CBCcatalog}). Furthermore, on August 17, 2017, at 12:41:04 UTC a binary neutron star (BNS) inspiral signal (GW170817) was observed \citep{PhysRevLett.119.161101}. Less than two seconds later, the short gamma-ray burst (sGRB) GRB 170817A was detected by two space-based instruments: the Gamma-ray Burst Monitor (GBM) onboard \emph{Fermi} \citep{2017ApJ...848L..14G}, and the spectrometer anti-coincidence shield (SPI-ACS) onboard \emph{INTEGRAL} \citep{2017ApJ...848L..15S}. This joint observation provided the first direct evidence that at least a fraction of sGRBs have a BNS system as progenitor, as predicted by \citealt{1989Natur.340..126E, 1986ApJ...308L..43P, 1991AcA....41..257P}. Short GRBs are typically expected to result in a long lasting, multi-wavelength afterglow emission in X-ray, optical, and radio bands (for a review see \citealt{2007PhR...442..166N, 2014ARA&A..52...43B, 2015JHEAp...7...73D}).

The extensive electromagnetic (EM) observational campaign using the well-constrained, three-detector skymap from the detection of GW170817 led to the discovery of an optical transient (SSS17a/AT 2017gfo) in the host galaxy NGC 4993 \citep{Coulter2017}; the counterpart was also detected in ultraviolet and infrared wavelengths \citep{2017ApJ...848L..12A}. Photometric and spectroscopic observations of the counterpart support the hypothesis that BNS mergers are sites of r-process nucleosynthesis of heavy elements that decay, thus powering so-called kilonova emission in UV/optical/NIR (see, e.g., \citealp{1998ApJ...507L..59L, 2005astro.ph.10256K, 2016AdAst2016E...8T, kilonovae, 2017Natur.551...80K, 2017ApJ...851L..21V, 2017Sci...358.1565E, 2017Natur.551...67P}). Several days after the BNS merger, X-ray \citep{2017Natur.551...71T} and radio \citep{2017Sci...358.1579H} emissions were also discovered at the transient's position (see also \citealt{2017ApJ...848L..12A} and references therein). These observations are consistent with the expected interaction of merger ejecta with the interstellar medium on timescales up to years (see, e.g., \citealt{2011Natur.478...82N, 2015MNRAS.450.1430H, 2016ApJ...831..190H}). Data from exhaustive followup in X-ray, radio and optical covering almost one year allowed detailed modeling of emission mechanisms, such as an off-axis structured jet (see, e.g., \citealt{2018arXiv180106164D,2018arXiv180306853D,2018arXiv180103531M,2018Natur.554..207M,2018ApJ...853L...4R}). The degeneracy among the various models has been broken with late-time radio observations that support the emergence of a relativistic jet from the BNS merger \citep{2018arXiv180800469G,2018arXiv180609693M}.

Besides compact binary mergers, other transient GW sources that may be observed by ground-based interferometers include the core-collapse of massive stars, which are expected to emit GWs if some asymmetry is present (see \citealt{2006RPPh...69..971K, 2009CQGra..26f3001O, 2015APS..APR.C7002G} for an overview). The core-collapse of a massive star is accompanied by supernova (SN) emissions, starting in the ultraviolet and soft X-ray bands from the shock breakout of the stellar surface (see, e.g., \citealp{1977ApJS...33..515F,1978ApJ...223L.109K,2016A&A...587A.147A,1992ApJ...393..742E}), and followed by emissions at optical and radio frequencies that typically start from days to weeks after the collapse and last for weeks up to years. Neutrinos are also emitted during core-collapse supernovae as confirmed on February 23, 1987 when MeV neutrinos were detected from SN 1987A in the Large Magellanic Cloud (at a distance of $\sim$50 kpc) by the Kamiokande-II \citep{1987PhRvL..58.1490H} and the Irvine-Michigan-Brookhaven \citep{1987PhRvL..58.1494B} neutrino detectors, a few hours before its optical counterpart was discovered. In addition, GRBs and SNe are expected to produce relativistic outflows in which particles (protons and nuclei) can be accelerated and produce high-energy neutrinos by interacting with the surrounding medium and radiation (see, e.g., \citealt{2018PhRvD..97h1301M}).

One further class of transient GW sources are magnetars, i.e., rotating NSs with very intense magnetic fields ($\sim 10^{15}$ G). Theoretical models predict that when these stars undergo starquakes, asymmetric strains can temporarily alter the geometry of the star and GWs could be emitted (see, e.g., \citealt{2011PhRvD..83j4014C}). Electromagnetic phenomena possibly associated with magnetar starquakes include Soft Gamma Repeaters (SGRs) and Anomalous X-ray Pulsars (AXPs), sources that sporadically emit short bursts of gamma-rays and X-rays (see \citealt{2008A&ARv..15..225M} for a review). Starquakes can also cause radio/X-ray pulsar glitches: sudden increases in the rotational frequency of a highly magnetized, rotating NS (pulsar) followed by exponential decays, which bring the pulsar rotational frequency back down to its initial value (see, e.g., \citealt{2011MNRAS.414.1679E}).

During O1 and O2, extensive EM observing campaigns searched for counterparts to GW candidates identified in low-latency. Significant improvements were made between these two observing runs regarding the data analysis software and source modeling, allowing important additional information to be distributed in low-latency during O2. For CBC events, 3D sky localization maps were released, providing information about the direction and the luminosity distance of the source \citep{2041-8205-829-1-L15}, while in O1, only 2D sky localization maps were provided, without distance information. During O2, probabilities that at least one low-mass object was present in the coalescing binary system and that tidally disrupted material formed a massive accretion disk around the merged object were reported. This information is useful in assessing the likelihood that a merger could power an EM transient \citep{2012PhRvD..86l4007F, 2014ApJ...791L...7P}.

During O1 and the first part of O2, with the GW network formed only by the two Advanced LIGO interferometers, sources were typically localized in sky areas ranging from a few hundreds to several thousands of square degrees (see, e.g., \citealt{2016ApJ...826L..13A, 2016PhRvL.116x1103A, 2017ApJ...851L..35A}). Improvements in localization areas were made since Advanced Virgo joined the gravitational-wave detector network starting August 1, 2017. For instance, GW170814 and GW170817 were localized by the three detector network within a few tens of square degrees, see \citealt{2017PhRvL.119n1101A,PhysRevLett.119.161101}. 

Jointly observing the same event in both gravitational waves and electromagnetic radiation provides complementary insights into the progenitor and its local environment. The GW signal is key to determining several physical properties of the source such as the masses and system properties (inclination, orientation, spin, etc.). The EM counterpart provides information about radioactive decay, shocks, the emission mechanism of the central engine, magnetic fields, beaming and also probes the surrounding environment of the source (see for instance \citealt{2014ARA&A..52...43B}). The detection of an EM counterpart also can give precise localization and lead to the identification of the host galaxy of the source. The distance estimated from the GW data combined with the measured redshift of the host galaxy enables measurement of the Hubble constant \citep{1986Natur.323..310S, 2005ApJ...629...15H, 2010ApJ...725..496N, 2013arXiv1307.2638N, 2017Nature..24471, 2018MNRAS.475.4133S, 2018arXiv180610596H,2018PhRvL.121b1303V}. Precise measurements of the host galaxy distance and the binary inclination given by the EM observations can be used to reduce the degeneracy in the GW parameter estimation (see, e.g., \citealt{2017ApJ...851L..36G, 2018ApJ...854L..31C, 2018ApJ...853L..12M, 2018arXiv180705226C}). Furthermore, the detection of an EM counterpart may increase the confidence in the astrophysical origin of a weak GW signal \citep{1993ApJ...417L..17K}. It also provides constraints on the relative merger rates of the two classes of compact binaries (BNS and NS-BH), on the beaming angle of sGRBs, and the NS equation of state \citep{2010CQGra..27q3001A, 2013PhRvL.111r1101C, 2014ApJ...791L...7P, 2015ApJ...809...53C, 2015ApJ...806..263D, 2015ApJ...799...69R, 2016arXiv160603043S, 2018ApJ...852L..29R}. Finally, joint GW and EM observations can provide constraints on fundamental physics \citep{2017ApJ...848L..13A}. 

\smallbreak

In this paper, we describe the identification of GW transients and the distribution of GW alerts performed during O2 by the LIGO and Virgo collaborations. We also detail the GW event information shared with the astronomy community and give an overview of the EM follow-up strategies. 

In section~\ref{sec:GW_pipelines} we present an overview of the online GW analysis, with a description of the online analysis detection pipelines, the vetting and the approval processes for potential GW events. In section~\ref{sec:Alerts} we summarize the GW alerts that were distributed during O2 and the properties of the gravitational wave candidates after the offline analysis. We describe the information that was shared with astronomers including how this was used during the electromagnetic/neutrino follow-up activities. Finally, in section~\ref{sec:conclu} we present our conclusions.

\section{Online Gravitational Wave Analysis}
\label{sec:GW_pipelines}
In this section, we describe the two classes of searches for GW transients, modeled and unmodeled, that contributed triggers for low-latency EM follow-up (in section~\ref{onlinepipeline}). We also present the full vetting and validation process of candidate events (in section~\ref{Validationprocess}) and distribution of low-latency alerts during O2 (in section~\ref{O2triggers}). Offline search pipelines\footnote{The offline O2 results were obtained after a complete regeneration of O2 strain data with noise subtraction performed for the LIGO detectors \citep{Davis:2018yrz}.} also led to the identification of additional candidate events GW170729 and GW170818 (see \citealt{CBCcatalog}).

\subsection{Brief description of online pipelines}
\label{onlinepipeline}

The modeled (CBC) searches specifically look for signals from compact binary mergers of neutron stars and black holes (BNS, NS-BH, and BBH systems). The unmodeled (burst) searches on the other hand, are capable of detecting signals from a wide variety of astrophysical sources in addition to compact binary mergers: core-collapse of massive stars, magnetar starquakes, and more speculative sources such as intersecting cosmic strings or as-yet unknown GW sources.

\subsubsection{Online modeled searches}
\label{modeledSearches}

GstLAL \citep{PhysRevD.95.042001}, MBTAOnline (Multi-Band Template Analysis, \citealt{0264-9381-33-17-175012}) and PyCBC Live \citep{PyCBCLive} are analysis pipelines designed to detect and report compact binary merger events with sub-minute latencies.
Such pipelines use discrete banks of waveform templates to cover the target parameter space of compact binaries and perform matched filtering on the data using those templates, similar to the offline analyses \citep{Usman:2015kfa, PhysRevD.95.042001} that produced the O1 and O2 catalog of compact binaries \citep{CBCcatalog}.
The online and offline analyses differ in various ways. The most important configuration choices of online analyses are reviewed here.

\begin{deluxetable*}{r|c|c|c}[t]
\tablecolumns{4}
\tabletypesize{\scriptsize}
\tablecaption{Major parameters of O2 online search pipelines based on compact binary merger waveform models.}
\tablehead{
\hspace{0.5cm} & PyCBC Live & GstLAL & MBTAOnline}
\startdata
Total mass & $2\msun$ to $500\msun$\tablenotemark{a} & $2\msun$ to $150\msun$\tablenotemark{a} & $2\msun$ to $100\msun$ \\
\hline
Mass ratio & $1$ to $98$ & $1$ to 98 & $1$ to $99$ \\
\hline
Minimum component mass & $1\msun$ & $1\msun$ & $1\msun$ \\
\hline
Spin magnitude ($m<2\msun$) & 0 to $0.05$ & 0 to $0.05$ & 0 to $0.05$ \\
\hline
Spin magnitude ($m>2\msun$) & 0 to $0.998$ & 0 to $0.999$ & 0 to $0.9899$ \\
\hline
Single-detector SNR threshold for triggering & $5.5$ & $4$\tablenotemark{b} & $5.5$\tablenotemark{c} \\
\enddata
\tablenotetext{\rm{a}}{The maximum total mass for PyCBC Live and GstLAL is in fact a function of mass ratio and component spins (\citealt{DalCanton:2017ala, Mukherjee:2018}) and we indicate the highest total mass limit over all mass ratios and spins. The offline GstLAL search uses a template bank extended to a larger maximum total mass of 400$\msun$.}
\tablenotetext{\rm{b}}{This threshold was applied to the two LIGO detectors only for the online GstLAL analysis. The minimum trigger SNR in Virgo was not determined by an explicit threshold, but instead by a restriction to record at most 1 trigger per second in a given template.}
\tablenotetext{\rm{c}}{MBTAOnline uses a higher LIGO SNR threshold ($6$) to form coincidences with Virgo.}
\label{table:pipeline_params}
\end{deluxetable*}
The mass and spin parameter space considered by the online pipelines in O2 is summarized in Table \ref{table:pipeline_params}. All pipelines assume that, while the gravitational wave signal dwells in the detector sensitive band, the spins of the compact objects are aligned or antialigned with the orbital angular momentum, and that orbital eccentricity is negligible.
Additional details of the PyCBC Live and GstLAL banks can be found in \citealt{DalCanton:2017ala} and \citealt{Mukherjee:2018}, respectively.
In the case of PyCBC Live, the online and final offline analyses covered exactly the same space.
For GstLAL, the offline bank extended to a larger total mass of $400\msun$.

A matched-filtering analysis is performed by each pipeline, producing triggers for each detector's data stream whenever the matched-filter single-detector signal to noise ratio (SNR) peaks above a threshold given in Table \ref{table:pipeline_params}.
Due to the small probability of a signal being detectable in Virgo and not in LIGO during O2, PyCBC Live did not use Virgo to produce triggers; Virgo's data was nevertheless still analyzed and used for the sky localization of candidates from LIGO.

Matched filtering alone is insufficient in non-Gaussian detector noise, producing frequent non-astrophysical triggers with large SNR \citep{TheLIGOScientific:2016zmo}.
Pipelines can choose among different techniques to mitigate this effect: calculating additional statistics based on the template waveform (signal-based vetoes), explicitly zeroing out loud and short instrumental transients before matched-filtering (gating), and vetoing triggers based on known data-quality issues which are reported with the same latency as the strain data itself.
In O2, all matched-filter searches employed signal-based vetoes; PyCBC Live and MBTAOnline applied vetoes based on low-latency data-quality information, while GstLAL applied gating.

The trigger lists produced by matched filtering and cleaned via the aforementioned procedures are searched for coincidences between detectors.
Coincident triggers are ranked based on their SNRs and signal-based vetoes and the consistency of their SNRs, arrival times and arrival phases at the different detectors with an astrophysical signal.
The pipelines construct this ranking and convert it to a statistical significance in different ways, described next.
A measure of significance produced by all pipelines for each candidate is the estimated false-alarm rate (FAR), i.e., the rate at which we expect events with at least as high a ranking as the candidate to be generated due to detector noise.

MBTAOnline constructs a background distribution of the ranking statistic by making every possible coincidence from single-detector triggers over a few hours of recent data.
It then folds in the probability of a pair of triggers passing the time coincidence test.

PyCBC Live's ranking of coincident triggers in O2 was somewhat simpler than the final offline analysis: it did not account for the variation of background over the parameter space \citep{0004-637X-849-2-118} and it did not include the sine-Gaussian signal-based veto \citep{Nitz:2017lco}.
PyCBC Live estimated the background of accidental coincidences by using time shifts between triggers from different detectors, as done by the offline analysis \citep{Usman:2015kfa}.
The amount of live time used for background estimation in PyCBC Live was 5 hours, to be compared with $\sim$5 days of the offline analysis.
This choice limited the inverse false-alarm rate of online detections to $\sim$100 years maximum, insufficient for claiming a very significant detection, but adequate for generating rapid alerts for astronomers.
On the other hand, this choice gave the background estimation a faster response to variations in noise characteristics, which is useful considering the limited data quality flags available to the online analysis.

GstLAL calculates the significance of triggers by constructing a likelihood-ratio ranking statistic that models the distribution of trigger properties for noise and GW events \citep{Cannon:2015gha}.
The background is computed by synthesizing likelihood ratios from a random sampling of a probability density that is estimated using non-coincident triggers accumulated over the course of an observing run, which are taken to be noise.

\subsubsection{Online unmodeled searches}
\label{unmodeledSearches}

The two unmodeled  signal searches (burst), cWB and oLIB, work by looking for excess power in the time-frequency (TF) domain of the GW strain data \citep{PhysRevD.93.042004, PhysRevD.95.104046}. The cWB pipeline does this by creating TF maps at multiple resolutions across the GW detector network
and identifying clusters of TF data samples with power above the baseline detector noise.
Excess power clusters in different detectors that overlap in time and frequency indicate
the presence of a GW event. The signal waveforms and the source sky location are reconstructed
with the maximum likelihood method by maximizing over all possible time-of-flight delays in the
detector network. The cWB detection statistic is based on the coherent energy obtained
by cross-correlating the signal waveforms reconstructed in the detectors. It is compared to
the corresponding background distribution to identify significant GW candidates.

oLIB uses the $Q$ transform to decompose GW strain data into several TF planes of constant quality factors $Q$, where $Q\sim \tau f_0$. Here, $\tau$ and $f_0$ are the time resolution and central frequency of the transform's filter/wavelet,  respectively. The pipeline flags data segments containing excess power and searches for clusters of these segments with identical $f_0$ and $Q$ spaced within 100 ms of each other. Coincidences among the detector network of clusters with a time-of-flight window up to 10 ms are then analyzed with a coherent (i.e., correlated across the detector network) signal model to identify possible GW candidate events.

Similarly to PyCBC Live, both cWB and oLIB use local time slides to estimate the background and calculate the candidates' false alarm rates (FAR is detailed in section~\ref{definitionFAR}).

\subsection{Vetting and approval process}
\label{Validationprocess}

During O1 and O2, all CBC and burst GW triggers were stored in an interactive database (see section~\ref{GraceDb}) and required to pass a series of vetting procedures, both automatic (section~\ref{Dataquality}) and manual (section~\ref{Humanvetting}), with the help of supervised protocols (\ref{supervisedEM}).

\subsubsection{GraceDb and LVAlert}
\label{GraceDb}

The Gravitational-wave Candidate Event Database (GraceDb\footnote{https://gracedb.ligo.org}) is a centralized hub for aggregating and disseminating information about candidate events from GW searches.
It features a web interface for displaying event information in a human-friendly format, as well as a representational state transfer application program interface (RESTful API) for programmatic interaction with the service.
A Python-based client code package is also maintained to facilitate interactions with the API; this set of tools allows users to add new candidate events to the database, annotate existing events, search for events, upload files, and more.

During O2, GraceDb sent push notifications about candidate event creation and annotation to registered listeners via the LIGO/Virgo Alert System (LVAlert), a real-time messaging service based on the Extensible Messaging and Presence Protocol (XMPP) and its publish-subscribe (pubsub) extension.
Python-based command-line tools were provided to send and receive notifications, create messaging nodes, and manage node subscriptions.
Typical receivers of LVAlert messages were automated follow-up processes which, when triggered, performed tasks such as parameter estimation or detector characterization for a candidate event (details in section~\ref{Dataquality}).

\subsubsection{Supervised follow-up process}
\label{supervisedEM}

Several follow-up processes responded to the entry of a GW candidate into GraceDb, notified by the arrival of an LVAlert message. Three of these processes were of immediate relevance to the EM follow-up effort: the low-latency sky localization probability map (skymap) generator for CBC triggers, BAYESTAR \citep{PhysRevD.93.024013}, the tracker of candidate event status/incoming information and alert generator/sender (\textit{approval\_processorMP}), and the tracker of other follow-up processing (\textit{eventSupervisor}).

In particular, \textit{approval\_processorMP} was responsible for the decision to send alerts based on the following incoming state information: basic trigger properties from the pipelines (FAR, event time, detectors involved with the trigger), data quality and data products (sections~\ref{Dataquality} and~\ref{definitionclassificationCBC}), detector operator and advocate signoffs determining the result of human vetting (section~\ref{Humanvetting}), and other labels identifying time-correlated external triggers or signal injections performed in hardware at the sites. Hardware signal injections are simulated GW signals created by physically displacing the detectors' test masses \citep{PhysRevD.95.062002}. Triggers with false alarm rates below an agreed-upon FAR threshold and with no injection or data quality veto labels generated alerts to the astronomers involved in the LIGO and Virgo EM follow-up community via the GCN network (see section~\ref{GCNNetwork}) and GraceDb web services.

\subsubsection{Online automatic data vetting}
\label{Dataquality}

State information was provided to low-latency analysis pipelines indicating when the detectors' data were suitable for use in astrophysical analysis. This included times when the detectors were operating in a nominal state and data calibration was accurate. The low-latency pipelines also dealt with the additional challenge of transient noise artifacts known as glitches, which often occurred in the detectors' data \citep{TheLIGOScientific:2016zmo,VirgoO2Paper}.

To reduce the effect of glitches, which can mimic true GW signals to some degree but are uncorrelated in the GW detectors, multiple strategies were employed by LIGO and Virgo, including automatically produced data quality vetoes and human vetting of candidate events (see section~\ref{Humanvetting}). Data quality vetoes indicated times when a noise source known to contaminate the astrophysical searches was active. These vetoes were defined using sensors that measured the behavior of the instruments and their environment. This data quality information was applied in several steps. A set of data quality vetoes were generated in real time and provided to the low-latency pipelines alongside the detector state information. If a candidate GW event occurred during a time that had been vetoed, it was not reported for EM follow-up. Given that these vetoes could potentially prevent a true GW signal from being distributed, this category of data quality information was reserved for severe noise sources.

In parallel with this effort, low-latency algorithms searched LIGO data for correlations between witness sensors and the GW strain data to identify noise sources that might not have been included in defined data quality vetoes. For example, iDQ (a streaming machine learning pipeline based on \citealt{2013CQGra..30o5010E} and \citealt{PhysRevD.88.062003}) reported the probability that there was a glitch in h(t) based on the presence of glitches in witness sensors at the time of the event. In O2, iDQ was used to vet unmodeled low-latency pipeline triggers automatically.

\subsubsection{Human vetting}
\label{Humanvetting}

Human vetting of GW triggers was a critical part of the EM follow-up program, and had to be completed before sending any alert to the astronomers during O2. Potentially interesting triggers were labeled by \textit{approval\_processorMP} to require signoffs from follow-up advocates and operators at each relevant detector site.

Different groups of persons from the collaborations were involved in the decision-making process including Rapid Response Teams (RRT) with commissioning, computing, and calibration experts from each of the detector sites, pipeline experts, detector characterization experts, and EM follow-up advocates.

First, the on-site operators had to check the status of the instruments within one minute of the trigger, to ensure that unusual events (thunderstorms, trucks driving close to the buildings, etc.) did not happen at the time of the GW trigger and that the interferometer status was nominal.

Second, the experts and on-duty advocates met during an on-call validation process organized immediately after being notified of the trigger. All previously mentioned algorithms in section \ref{Dataquality} and additional data quality information not accessible at low-latency timescales were considered. For example, the Omega scan and Omicron scan algorithms \citep{Chatterji:2004qg, omicron} created time-frequency visualizations of witness sensor data around the time of the candidate event. This allowed for detailed views of instrumental or environmental noise that could potentially influence the detectors' GW strain data. In O2, this information was used to identify false triggers due to noise and veto them before they were reported for EM follow-up.

Third, pipeline experts were asked to check pipeline results, in particular to evaluate the significance of marginal triggers.
In the case of more than one viable candidate event (within 1 second for CBC and 5 seconds for burst triggers), the advocates selected the most promising candidate based on pre-established criteria (e.g. lowest FAR, choosing CBC over burst triggers). 

When Virgo joined the LIGO network, it was not used to estimate the FAR of the GW candidate, but only to constrain the sky localization.

Finally, the EM follow-up advocates selected the skymap to send depending on the cross-checks done by the RRT at the different instrument sites, with priority given to the two LIGO detectors as the most sensitive
instruments in the network. Then, they released the skymap to the external community and composed the GCN circular (see section~\ref{GCNNetwork}).

When necessary, the pipeline experts and the data quality team, with the help of the RRT, recommended a retraction after days or weeks, 
using extended data investigation \citep{VirgoO2Paper} and/or updated FAR calculation based on additional background data.

\subsection{Triggers distributed during O2}
\label{O2triggers}

During O2, only GW candidates that passed the above series of checks were distributed to partner astronomers (see Table~\ref{tab:characdistributed triggers}). Approximately half the triggers were rejected during the human vetting process (described in section~\ref{Humanvetting}) due to the presence of glitches (see \citealt{O1DetcharCBC, VirgoO2Paper}). The number of vetoed triggers decreased by 80\% from the first half of the O2 observing run to the second due to pipeline software upgrades to avoid transient noise.
Other candidate events were vetoed because they were generated due to specific hardware problems or by not meeting the requirements of the O2 alert distribution policy (e.g., single-detector triggers, FAR being above the O2 threshold value, and very high latencies).

\smallbreak

The list of distributed triggers during O2 and their online characteristics are provided in Table~\ref{tab:characdistributed triggers}. We note that both CBC and burst pipelines identified the loudest GW candidate events. Six low-latency CBC candidates were ultimately confirmed as GW detections and are described in detail in \citet{CBCcatalog}. G288732 (subsequently named GW170608) occurred when LIGO-Hanford was undergoing angle-to-length decoupling (a regular maintenance procedure which minimises the coupling between the angular position of the test-mass mirrors and the measurement of the strain) \citet{2017ApJ...851L..35A},  whereas for G268556 (i.e., GW170104), the calibration state was not nominal \citep{2017PhRvL.118v1101A}, creating a high latency in the distribution of alerts. These two real events were recovered due to expert vigilance rather than automated procedures. Moreover, G298048 (i.e., GW170817) was first identified as a single-detector trigger in the LIGO-Hanford data; a glitch in LIGO-Livingston caused the trigger to be rejected and the SNR was too low in Virgo to be detected.

Burst event G270580 was retracted offline due to its correlation with seismic noise \citep{GCN21281}. The CBC candidates G275697, G275404 and G299232 were not present in the offline pipeline analysis whereas other marginal candidates have been listed in \citet{CBCcatalog}. The burst triggers G274296, G277583, G284239 and G298389 are consistent with background noise based on their event parameters and FARs; hence they are of no further interest. The high latency in sending alerts for the two oLIB events, G284239 and G298389, was due to their skymap generation. 

\begin{deluxetable*}{c|c|c|c|cc|cc}[t]
\tablecolumns{8}
\tabletypesize{\scriptsize}
\tablewidth{1.0\columnwidth} 
\tablecaption{Characteristics of the distributed triggers which passed the EM follow-up validation process: time of the GW candidate event, status of the event after offline analysis (Confident i.e., Confidently detected GW event; Retracted due to further noise investigation; or NFI i.e., No Further Interest, not present in the offline analysis or consistent with noise), nature of the candidate with an EM-Bright classifier (described in section~\ref{definitionclassificationCBC}; if N/A, classifier not available for burst triggers), list of online searches that detected the candidate event (the pipeline selected for the distributed alert is in bold, some of the pipelines are not indicated in the first GCN circular since they reported the trigger with larger latency), its FAR (online pipeline-dependent, see section~\ref{definitionFAR}), the latency between the event time and event submission into GraceDb and the delay between the event time and the alert distribution (first GW notice sent to GCN for distribution to partners, see section \ref{GCNNetwork}).}
\tablehead{
\multirow{2}{*}{GW ID} & Event Time & \multirow{2}{*}{Final status} & Source & \multicolumn{2}{c|}{Triggers} & \multicolumn{2}{c}{Latency (min)}  \\	
& UTC & & Classification & Search(es)\tablenotemark{a} & \small{ Online FAR (yr$^{-1}$)} & GraceDb submission & Initial GCN notice  
}
\startdata
 G268556 & \multirow{2}{*}{2017-01-04 10:11:58} & \multirow{2}{*}{Confident} & BBH & \textbf{PyCBC}, cWB & \multirow{2}{*}{1.9 }\tablenotemark{b} & \multirow{2}{*}{264} & \multirow{2}{*}{395}  \\
 GW170104 & & & EM-Bright: 0$\%$ & GstLAL, oLIB &  &  &\\
\hline
\multirow{2}{*}{G270580} & \multirow{2}{*}{2017-01-20 12:30:59} & \multirow{2}{*}{Retracted} & Burst & \multirow{2}{*}{\textbf{cWB}}  & \multirow{2}{*}{5.0} & \multirow{2}{*}{2} & \multirow{2}{*}{67}  \\
 & & & EM-Bright: N/A & & & &  \\
\hline
\multirow{2}{*}{G274296} & \multirow{2}{*}{2017-02-17 06:05:53} & \multirow{2}{*}{NFI} & Burst & \multirow{2}{*}{\textbf{cWB}}  & \multirow{2}{*}{5.4} & \multirow{2}{*}{715} & \multirow{2}{*}{813} \\
 & & & EM-Bright: N/A & & & & \\
\hline
 \multirow{2}{*}{G275404} & \multirow{2}{*}{2017-02-25 18:30:21} & \multirow{2}{*}{NFI} & NS-BH & \multirow{2}{*}{\textbf{PyCBC}, GstLAL}  & \multirow{2}{*}{6.0} & \multirow{2}{*}{$<$1} & \multirow{2}{*}{24} \\
   & & & EM-Bright: 90$\%$ & & & &  \\
 \hline
 \multirow{2}{*}{G275697} & \multirow{2}{*}{2017-02-27 18:57:31}& \multirow{2}{*}{NFI} & BNS & \textbf{PyCBC}, GstLAL, & \multirow{2}{*}{4.5} & \multirow{2}{*}{$<$1} & \multirow{2}{*}{27}  \\
   & & & EM-Bright: 100$\%$ & MBTAOnline & & &  \\
 \hline
 \multirow{2}{*}{G277583} & \multirow{2}{*}{2017-03-13 22:40:09} & \multirow{2}{*}{NFI} & Burst & \multirow{2}{*}{\textbf{cWB}} & \multirow{2}{*}{2.7} & \multirow{2}{*}{3} & \multirow{2}{*}{30}  \\
    & & & EM-Bright: N/A & & & & \\
 \hline
 \multirow{2}{*}{G284239} & \multirow{2}{*}{2017-05-02 22:26:07} & \multirow{2}{*}{NFI} & Burst & \multirow{2}{*}{\textbf{oLIB}} & \multirow{2}{*}{4.0} & \multirow{2}{*}{12} & \multirow{2}{*}{963}  \\
    & & & EM-Bright: N/A & & & & \\
 \hline
 G288732 & \multirow{2}{*}{2017-06-08 02:01:16} & \multirow{2}{*}{Confident} & BBH & \multirow{2}{*}{\textbf{PyCBC}, cWB, GstLAL\tablenotemark{c}}  & \multirow{2}{*}{2.6}\tablenotemark{b} & \multirow{2}{*}{650} & \multirow{2}{*}{818}  \\
  GW170608 & & & EM-Bright: 0$\%$ & & & & \\
  \hline
  G296853 & \multirow{2}{*}{2017-08-09 08:28:22} & \multirow{2}{*}{Confident} & BBH & \textbf{GstLAL}, cWB & \multirow{2}{*}{0.2} & \multirow{2}{*}{$<$1} & \multirow{2}{*}{49} \\
  GW170809 & & & EM-Bright: 0$\%$ & MBTAOnline & & & \\
  \hline
  G297595 & \multirow{2}{*}{2017-08-14 10:30:44} & \multirow{2}{*}{Confident} & BBH & \textbf{GstLAL}, oLIB,  & \multirow{2}{*}{1.2$\times10^{-5}$} & \multirow{2}{*}{$<$1} & \multirow{2}{*}{31} \\
  GW170814 & & & EM-Bright: 0$\%$ & PyCBC, cWB\tablenotemark{d} & & & \\
  \hline
  G298048 & \multirow{2}{*}{2017-08-17 12:41:04} & \multirow{2}{*}{Confident} & BNS & \multirow{2}{*}{\textbf{GstLAL}\tablenotemark{d}, PyCBC}  & \multirow{2}{*}{1.1$\times10^{-4}$} & \multirow{2}{*}{6} & \multirow{2}{*}{27}\tablenotemark{e} \\
  GW170817 & & & EM-Bright: 100$\%$ & & & & \\
  \hline
  \multirow{2}{*}{G298389} & \multirow{2}{*}{2017-08-19 15:50:46} & \multirow{2}{*}{NFI} & Burst & \multirow{2}{*}{\textbf{oLIB}}  & \multirow{2}{*}{4.9} & \multirow{2}{*}{16} & \multirow{2}{*}{192} \\
    & & & EM-Bright: N/A & & & & \\
  \hline
  G298936  & \multirow{2}{*}{2017-08-23 13:13:59} & \multirow{2}{*}{Confident} & BBH & cWB, oLIB, \textbf{GstLAL},  &  \multirow{2}{*}{5.5$\times10^{-4}$} & \multirow{2}{*}{$<$1} & \multirow{2}{*}{22}  \\
  GW170823 & & & EM-Bright: 0$\%$ & PyCBC, MBTAOnline & & & \\
  \hline
  \multirow{2}{*}{G299232} & \multirow{2}{*}{2017-08-25 13:13:37} & \multirow{2}{*}{NFI} & NS-BH & \multirow{2}{*}{\textbf{MBTAOnline}\tablenotemark{f}}  & \multirow{2}{*}{5.3} & \multirow{2}{*}{$<$1} & \multirow{2}{*}{25} \\
    & & & EM-Bright: 100$\%$ & & & & \\
\enddata 
\tablenotetext{\rm{a}}{In bold, selected pipeline for distribution of the alert to O2 partners}
\tablenotetext{\rm{b}}{Due to the non-standard way in which this trigger was found, the PyCBC Live low-latency pipeline was run by hand over a short period of data (tens of minutes) in order to produce a trigger for follow-up as quickly as possible. The precision of the FAR estimate is limited by the use of a shorter than normal period of data.}
\tablenotetext{\rm{c}}{The online GstLAL trigger was identified as a single-detector trigger.}
\tablenotetext{\rm{d}}{The sky localization sent to partners five hours after the trigger time was derived from a PyCBC analysis after the high-amplitude glitch in LIGO-Livingston was windowed out \citep{GCN21513,PhysRevLett.119.161101}. The FAR was calculated with H1 only.}
\tablenotetext{\rm{e}}{The first circular sent to partners for G298048 informed that a GW candidate event with a single instrument was associated with the time of a Fermi GBM trigger \citep{GCN21505}. The initial Notice sent to partners, at 13:08 UTC, contained a skymap simply representing the quadrupolar antenna response function of the LIGO-Hanford detector over the entire sky.}
\tablenotetext{\rm{f}}{The sky localization sent to partners was derived from a PyCBC analysis \citep{GCN21693}}
\label{tab:characdistributed triggers}
\end{deluxetable*}

\section{Distribution of Alerts}
\label{sec:Alerts}
In this section, we present GW candidate information that we distributed and how this supported the EM observational campaigns.
\subsection{O2 partners network}

For O2, \LVC{} signed 95 memoranda of understanding (MoUs) with different institutions, agencies, and groups of astronomers from more than 20 countries. The goal was to enable multi-messenger observations of astrophysical events by GW detectors with a wide range of telescopes and instruments from EM and neutrino astronomy.

\smallbreak

During O2, 88 groups had operational facilities and the ability to receive and send notifcations regarding their observations through the GCN network (see section~\ref{GCNNetwork}). More than 100 space and ground-based instruments were involved in the EM follow-up campaign by covering radio, optical, near-infrared, X-ray and gamma-ray wavelengths. The telescopes sensitive in the optical bands were the most numerous, representing half of the instruments. The follow-up network also included three facilities capable of detecting high-energy neutrinos: IceCube, ANTARES, Pierre Auger, searching for transients in the northern and southern hemispheres.

\subsection{Distribution of the alerts via the GCN network}
\label{GCNNetwork}

The Gamma-ray Coordinates Network (GCN)\footnote{https://gcn.gsfc.nasa.gov} was adopted from the GRB community to be used as an alert platform for both LIGO/Virgo observations and multi-messenger follow-up. There are two types of GCN alerts: notices and circulars. During O2, the LIGO/Virgo GCN network was private; a requester had to be a member of the LV-EM Forum\footnote{https://gw-astronomy.org/} to receive and send any messages. This is in contrast with normal public operation of GCN as used by the GRB community for decades. The LV-EM Forum, which consists of a wiki and mailing list, allowed registered astronomers to access information about GW candidate events selected for follow-up observations.

GCN/LVC notices (i.e., LIGO/Virgo-astronomers notices) are machine-readable-computer-generated messages containing basic information about GW candidate events (e.g., time of the event and/or a sky localization probability map) or EM counterpart candidates.
For the LIGO and Virgo collaborations' alerts, three types of GCN/LVC notices were produced: \textit{preliminary}, \textit{initial}, and \textit{update}, although the preliminary notices were distributed only internally within the LIGO/Virgo collaborations while the others were sent to all members of the LV-EM forum. There was the possibility of sending a \textit{retraction} notice as well.
\begin{itemize}
\item The \textit{preliminary} notice contains only basic trigger information such as the trigger time (equivalent to the event UTC time), the online pipeline that generated the trigger, and the event false alarm rate. It may also contain a skymap if one is available.
It reports unvetted GW candidates and is produced $\sim$1-3 minutes after the actual event time. 
\item The \textit{initial} notice is available $\sim$20-1000 minutes after the event  (see Table~\ref{tab:characdistributed triggers}) and is the result of further processing and human vetting of the event (see section~\ref{Validationprocess}). In addition to the fields provided in the \textit{preliminary} notices, it contains a link to the first sky localization probability map and source classification information if available for CBC candidate events.
\item The \textit{update} notice is available from hours to months after the event and reports offline and parameter estimation analysis, in terms of improved FAR and sky localization.
\end{itemize}

During O2, there were 198 individuals and groups that received one or more of the three LVC notice types by any of the distribution methods and formats (i.e., VOevents, binary socket packets, and email-based methods).

\smallbreak

The GCN circulars are human-generated prose-style descriptions of the event or follow-up observations made. They are generally sent shortly after their associated notices. For example, the LVC team generated one or two circulars for each GW candidate event with the first being sent $\sim$1-2 hours after the event time. Circulars are largely used to give information about follow-up observations, characteristics of the instruments/telescopes, and EM counterpart candidates. During O2, there were 385 recipients of the LVC/astronomers private circulars.

\subsection{Information sent to observing partners}

During O1 and O2, LIGO/Virgo notices and circulars contained basic trigger information such as the event time, corresponding online pipeline name, the list of contributing instruments (H1, L1, V1), and a sky localization probability map. In the case of CBC triggers for O2, additional information about the nature of the source (see section~\ref{definitionclassificationCBC}) and its localization with distance (see section~\ref{localizationO2}) was provided.

\begin{deluxetable*}{c|c|cc|cc|c|c|c}[t]
\tablecolumns{9}
\tabletypesize{\scriptsize}
\tablewidth{1.0\columnwidth}
\tablecaption{Properties of the GW alerts including the network SNR, the candidate event FAR, the sky localization area, and luminosity distances. Note that sky localization area and luminosity distances for the online search can differ from results mentioned in the distributed GCNs\tablenotemark{a}. Furthermore, the distance estimates stated in GCN circulars are the a posteriori mean $\pm$ standard deviation while distance estimates and confidence intervals stated in the table are the a posteriori median and central 90$\%$ intervals. For confident GW events, the table shows results obtained from the offline refined analysis for comparison. Network SNR and FAR are the offline analysis results obtained with the pipeline selected for online distribution of the alerts (see Table~\ref{tab:characdistributed triggers} and Table~1 in \citet{CBCcatalog}). Offline luminosity distance and sky localization area (50\% and 90\% confidence regions) are listed also in Table~8 from \citet{CBCcatalog}, with small differences arising from using the posterior data samples directly versus the generated skymap files.
A: \citet{GCN20364} - B: \citet{CBCcatalog} - C: \citet{GCN20486} - D: \citet{GCN20520, GCN21281} - E: \citet{GCN20689} - F: \citet{GCN21284} - G: \citet{GCN20738} - H: \citet{GCN20840,GCN20982} - I: \citet{GCN20763} - J: \citet{GCN20833,GCN20983} - K: \citet{GCN20860} - L: \citet{GCN21060} - M: \citet{GCN21221} - N: \citet{GCN21431}  - O: \citet{GCN21474} - P: \citet{GCN21510} - Q: \citet{GCN21513} - R: \citet{GCN21600} - S: \citet{GCN21656} - T: \citet{GCN21661} - U: \citet{GCN21693}}
\tablehead{\multirow{2}{*}{GW ID} & & \multicolumn{2}{c|}{Interferometers} & \multirow{2}{*}{SNR\tablenotemark{b}} & FAR\tablenotemark{b} & Luminosity distance & Sky localization area & \multirow{2}{*}{Ref} \\
 & & triggering & skymap (algorithm) &  & (yr$^{-1}$) & Median $\pm 90\%$ c.i. (Mpc) & 50\%/90\% c.i. ($\mathrm{deg}^2$) & }
\startdata
G268556 & online & \multirow{2}{*}{H1, L1} & H1, L1 (BAYESTAR) & 12.4  & 1.9 & $\mathrm{730^{+340}_{-320}}$  & 430/1630  & A \\
GW170104& offline & & H1, L1 (LALInference) &  13.0 & $<$ $1.4 \times 10^{-5}$ & $\mathrm{960^{+440}_{-420}}$ & 200/920  & B \\
\hline
\multirow{2}{*}{G270580} & online & \multirow{2}{*}{H1, L1} & \multirow{2}{*}{H1, L1 (LIB)} & 8.7 & 5.0  & \multirow{2}{*}{-} & 600/3120 & C \\
& offline & &  &  - & - &  & - & D \\
\hline
\multirow{2}{*}{G274296} & online &\multirow{2}{*}{H1, L1} & \multirow{2}{*}{H1, L1 (cWB)} & 11.0 &  5.4 & \multirow{2}{*}{-} & 430/2140  & E\\
& offline &  &  &  -  & - & & -  & F \\
\hline
\multirow{2}{*}{G275404} & online & \multirow{2}{*}{H1, L1} & H1, L1 (BAYESTAR) & 8.7 & 6.0 & $\mathrm{270^{+150}_{-130}}$ & 460/2100 & G \\
& offline & -\tablenotemark{c} & - & - & - & - & - & H \\
\hline
\multirow{2}{*}{G275697} & online & \multirow{2}{*}{H1, L1} & H1, L1 (BAYESTAR) & 8.7 & 4.5 & $180^{+90}_{-90}$ & 480/1820 & I \\
& offline & - \tablenotemark{c} & -  & - & - & - & - & J \\
\hline
\multirow{2}{*}{G277583} & online & \multirow{2}{*}{H1, L1} & \multirow{2}{*}{H1, L1 (cWB, LIB)\tablenotemark{d}} & 9.3 & 2.7 & \multirow{2}{*}{-} & 2130/12140\tablenotemark{d}  & K \\
& offline & &  & - & -  &  & - & -\\
\hline
\multirow{2}{*}{G284239} & online & \multirow{2}{*}{H1, L1} & \multirow{2}{*}{H1, L1 (LIB)} & 8.2 & 4.0 & \multirow{2}{*}{-} & 1030/3590 & L \\
& offline & &  & - & -  &  & - & - \\
\hline
G288732 & online & \multirow{2}{*}{H1, L1} & H1, L1 (BAYESTAR) & 12.7 & 2.6 & $\mathrm{310^{+200}_{-120}}$ & 230/860   & M \\
GW170608 & offline &  & H1, L1 (LALInference)  &  15.4 & $<$ $3.1\times 10^{-4}$ & $\mathrm{320^{+130}_{-100}}$  & 100/400 & B \\
\hline
G296853 & online & H1, L1 & H1, L1 (BAYESTAR) & 11.3 & 0.2 & $\mathrm{1080^{+520}_{-470}}$ & 320/1160 & N \\
GW170809 & offline & H1, L1 & H1, L1, V1 (LALInference) & 12.4 & $< 1.0\times 10^{-7}$ & $\mathrm{980^{+350}_{-350}}$ & 70/310\tablenotemark{e} & B \\
\hline
G297595 & online & \multirow{2}{*}{H1, L1} & H1, L1, V1 (BAYESTAR) & 16.1 & 1.2$\times10^{-5}$ & $\mathrm{480^{+190}_{-170}}$ & 22/97\tablenotemark{e} & O \\
GW170814 & offline &  & H1, L1, V1 (LALInference)  &  15.9 &  $< 1.0\times 10^{-7}$ & $\mathrm{570^{+190}_{-180}}$ & 16/87\tablenotemark{e} & B \\
\hline
\multirow{2}{*}{G298048} & \multirow{2}{*}{online} & H1 & H1 (BAYESTAR) & 14.5 & 1.1$\times 10^{-4}$ &  $40^{+20}_{-20}$ & 8060/24220 & P \\
 & & H1, L1 & H1, L1, V1 (BAYESTAR) & - & - & $40^{+10}_{-10}$ & 9/31\tablenotemark{e} & Q \\
GW170817 & offline & H1, L1 & H1, L1, V1 (LALInference) & 33.0 & $< 1.0\times 10^{-7}$ & $\mathrm{40^{+10}_{-10}}$ & 5/16\tablenotemark{e} & B \\
\hline
\multirow{2}{*}{G298389} & online & \multirow{2}{*}{H1, L1} & \multirow{2}{*}{H1, L1 (LIB)} & 15.6 & 4.9 & \multirow{2}{*}{-} & 250/800 & R  \\
& offline & &   & - &  - &  & - & - \\
\hline
\multirow{2}{*}{G298936} & \multirow{2}{*}{online} & H1, L1 & H1, L1 (BAYESTAR) & 11.3 &  5.5$\times10^{-4}$ & $\mathrm{1380^{+700}_{-670}}$ & 610/2140 & S \\
& & H1, L1 & H1, L1, V1 (BAYESTAR) & & - & $\mathrm{1540^{+690}_{-680}}$ & 277/1219\tablenotemark{e} & T \\
GW170823 & offline & H1, L1 & H1, L1 (LALInference) &  11.5 & $< 1.0\times 10^{-7}$ & $\mathrm{1860^{+920}_{-850}}$ & 440/1670 & B \\
\hline
\multirow{2}{*}{G299232} & online & H1, L1 & H1, L1, V1 (BAYESTAR) & 9.1 & 5.3 & $330^{+200}_{-160}$ & 451/2040\tablenotemark{e} & U \\
 & offline &   - \tablenotemark{c} & - &  -  & - & - & - & - \\
\enddata
\tablenotetext{\rm{a}}{Differences with respect to areas reported in GCN circulars are due to the rounding algorithm used to calculate the enclosed probability of 50\% and 90\%, which now is more accurate.}
\tablenotetext{\rm{b}}{The network SNR and the false alarm rate depend on the pipeline that triggered the event.}
\tablenotetext{\rm{c}}{No candidate event was found during the offline analysis.}
\tablenotetext{\rm{d}}{Localization is obtained as the arithmetic mean of cWB and LIB.}
\tablenotetext{\rm{e}}{All skymaps excluding those using Virgo data round sky localization areas to the nearest 10; otherwise, to the nearest 1.}
\label{tab:Distributedalerts}
\end{deluxetable*}

\subsubsection{Significance of the alerts}
\label{definitionFAR}
The significance of an online GW trigger during O2 was determined primarily by its false alarm rate (FAR). The FAR of a trigger quantifies the rate at which triggers of a given kind would be generated by an online detection pipeline from data that are void of any GW signal. Only triggers with a FAR below a pre-defined threshold were considered for EM follow-up. For the majority of O2, this threshold on FAR was once per two months ($1.9 \times 10^{-7}$ Hz). Thus, any trigger that was generated by an online detection pipeline which could have been generated simply by noise at a rate higher than once per two months was rejected for the EM follow-up program. The FAR estimation is specific to the pipeline that triggered the event (see section~\ref{onlinepipeline} and Table~\ref{tab:Distributedalerts} for distributed alerts and the associated FARs). In Table~\ref{tab:characdistributed triggers}, the eight triggers which were not confirmed as confident events were reported by five different pipelines. This is consistent with the expected number of false alarms from five pipelines in 118 days of coincident data from LIGO-Hanford, LIGO-Livingston, and Virgo \citep{CBCcatalog}.

\subsubsection{Source classification of CBC candidate events}
\label{definitionclassificationCBC}

\begin{figure}
    \centering
    \includegraphics[width=1.05\columnwidth, trim=1.5cm 1cm 1.5cm 1cm]{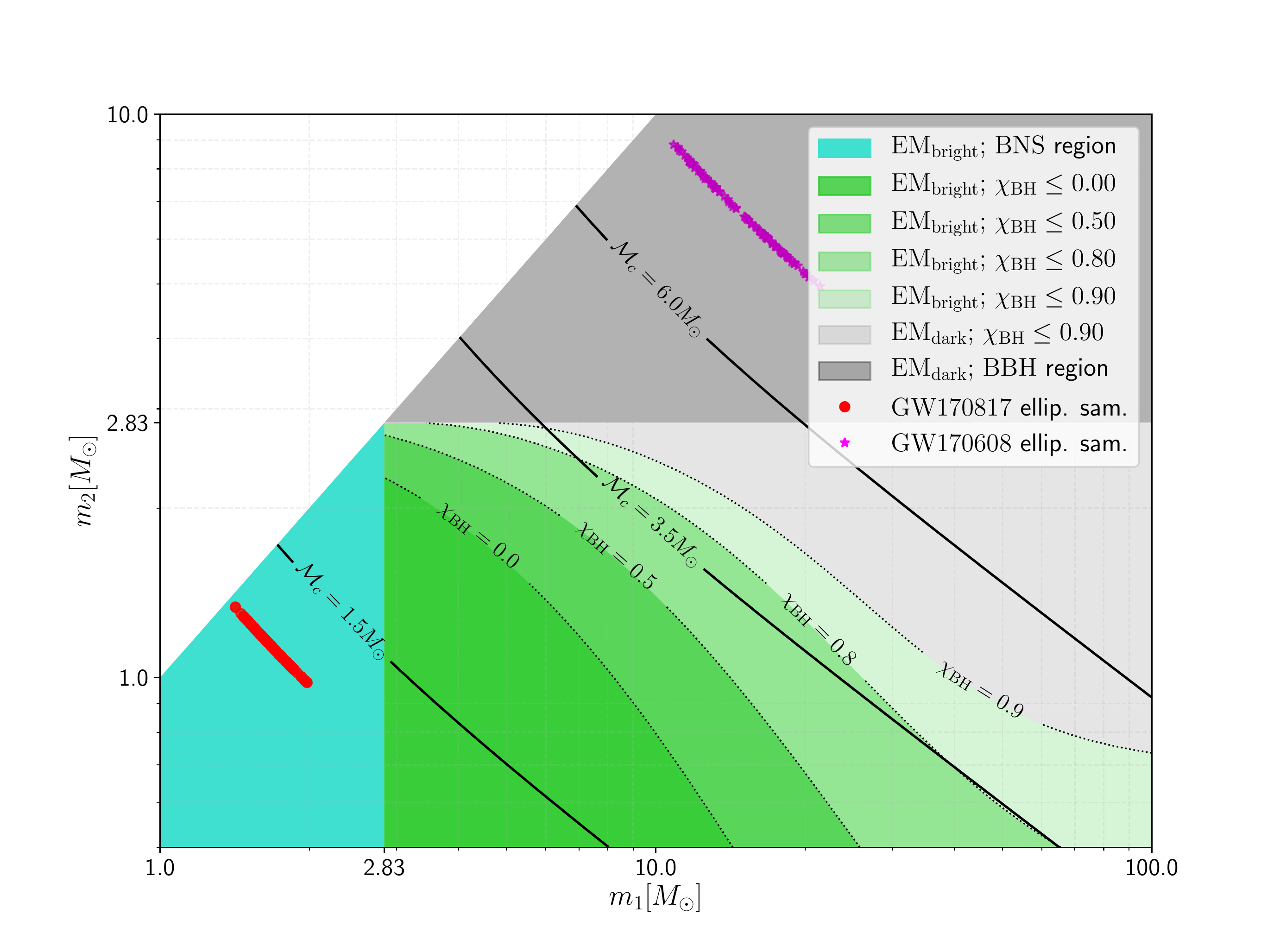}
    \caption{This figure shows the different regimes of operation by the source classifier used in O2.
    The BNS region is given by $(m_1, m_2) \in (0, 2.83) M_{\odot}$ (cyan). The upper limit of  
    $2.83 M_{\odot}$ is the maximum NS mass allowed by the 2H EOS \cite{Kyutoku:2010zd} used in the source
    classification software. Any system lying in this region is always considered to have
    an EM counterpart. The region $m_1 > 2.83 M_{\odot}$ with $m_2 \leq 2.83 M_{\odot}$ is the NS-BH 
    region of the source classifier. The baryonic mass left outside the BH is computed in this region. 
    In this scheme, the presence of any such matter is considered to provide the potential for an EM counterpart.
    This boundary of zero mass left outside the final BH is a function of the spin of the system. Increasing
    spin implies increased possibility of an EM counterpart. Dotted contours, tagged by different
    $\chi_{\text{BH}}$ values, indicate the EM-bright/dark boundary, with the bright region
	shaded for that particular $\chi_{\text{BH}}$ value.
	The chirp mass, $\mathcal{M}_c = (m_1m_2)^{3/5}/(m_1 + m_2)^{1/5}$,
	is the most accurately measured quantity from the GW waveform.
    A few chirp mass contours, $\mathcal{M}_c = (1.5, 3.5, 6.0) M_{\odot}$, are overlaid in solid lines 
    for comparison. Ellipsoid samples corresponding to a BBH merger, GW170608, and a BNS merger,
    GW170817 are also displayed showing consistent results with the source classifier. 
    }
\label{fig:EMBright}
\end{figure}

In an event where at least one of the component compact objects is a neutron star, the GW event is more likely to be accompanied by an EM counterpart. A new low-latency pipeline was implemented in O2 to provide observers with a source classification for compact binary coalescences. In low latency, the earliest event information that is available to use are the point estimates. The point estimates are values of the masses $(m_1, m_2)$, and the aligned components of spin $(\chi_1, \chi_2)$ of the template that triggered to give the lowest false alarm rate during the search. However, these point estimates have uncertainties and are expected to be offset with respect to the true component values \citep{Finn:1992xs, Jaranowski:1994xd, Cutler:1994ys, Poisson:1995ef, Arun:2004hn, Lindblom:2008cm, Nielsen:2012sb, Ohme:2013nsa, Hannam:2013uu}. Thus, any inferences drawn purely from the point estimates are prone to detection pipeline biases. To mitigate this effect, an effective Fisher formalism, introduced in \cite{Cho:2012ed}, was employed to construct an ellipsoidal region around the triggered point. This region, called the ambiguity ellipsoid, increases the chances of including the region of the parameter space that matches best with the true parameters of the source. This ambiguity ellipsoid is populated with 1000 points (a total of 1001 points including the original point estimate) which are called the ellipsoid samples. For each ellipsoid sample, the source classification quantities are computed. The dimensionality of the ambiguity ellipsoid is determined by the number of parameters required to compute the source classification quantities.

In O2 we delivered a twofold classification: the first which gives the probability that at least one neutron star is present in the binary, and the second which gives the probability that there is some baryonic mass left outside the merger remnant, i.e., the EM-Bright classification. While the first classification requires only one parameter for the inference to be conducted, namely the secondary mass component, the second classification, which is more model dependent, potentially needs more parameters than just the secondary mass. Indeed, we adopted the EM-Bright classification method from \cite{2012PhRvD..86l4007F} and implemented as in \cite{2014ApJ...791L...7P}, which uses three parameters $(m_1, m_2, \chi_1)$, the masses of the primary and secondary objects, and the aligned spin component of the primary object respectively. The method estimates the mass remaining outside the black hole after a NS-BH merger, which includes the mass of the accretion disk, the tidal tail, and/or unbound ejecta. A 3D ambiguity ellipsoid was generated around the triggered point to enclose a region of $90\%$ match within its boundary. This was done in the $(\mathcal{M}_c, \eta, \chi_1)$ parameter space where $\mathcal{M}_c = (m_1m_2)^{3/5}/(m_1 + m_2)^{1/5}$ is the chirp mass and $\eta = m_1m_2/(m_1 + m_2)^2$ is the symmetric mass ratio. This was achieved using infrastructure developed in \cite{Pankow:2015cra}. The fraction of ellipsoid samples with secondary mass less than 2.83 $M_{\odot}$ constituted the first classifier, namely, the probability that there is at least one neutron star in the binary.

Next, the mass left outside the black hole was computed for each ellipsoid sample using the fitting formula, Eq.(8) of \cite{2012PhRvD..86l4007F}. The fraction of ellipsoid samples for which this was greater than zero was calculated. This constituted the second classifier. 

It is important to note that Foucart's fitting formula is only valid in the NS-BH region of the parameter space. In O2, we made the assumption that binary neutron star mergers always emit EM radiation (i.e., EM-Bright: 100$\%$) and binary black hole mergers are always void of such emission (i.e., EM-Bright: 0$\%$, or equivalently, EM-Dark). Thus, the second classifier was computed only for ellipsoid samples with one component mass less than 2.83 $M_{\odot}$.

In Figure~\ref{fig:EMBright}, we can see different regions of the parameter space where the aligned spin component has been suppressed and only the mass values are shown. The blue region is the BNS parameter space where every ellipsoid sample is treated as EM-Bright. The dark grey region depicts the BBH parameter space where all ellipsoid samples are treated as EM-Dark. The light grey and green shaded regions are the NS-BH part of the parameter space where the EM-Bright probability is computed using Foucart's fitting formula. The green shaded regions show at which part of this parameter space the ellipsoid samples will give non-zero remnant mass outside the final black hole. The various shades of green discriminate between different $\chi_1$ values, so that, for example, an ellipsoid sample with mass values $(7.0, 2.0) M_{\odot}$ will give non-zero remnant mass outside the black hole according to Foucart's fitting formula if the value of $\chi_1$ is slightly greater than 0.5, but no mass left outside the black hole below this value. Additionally, this figure also shows ellipsoid samples for GW170817. The detection pipeline point estimate for this source was consistent with a binary neutron star system. Upon construction of the ambiguity ellipsoid around this point estimate, we found that all the ellipsoid samples lie completely within the blue shaded region that is always assumed to be EM-Bright. In contrast, a second event which is a binary black hole system, GW170608, is also depicted and its ellipsoid samples lie entirely in the EM-Dark regime. 

During O2, source classification information was provided (see Table~\ref{tab:characdistributed triggers}) on the basis of the detection pipeline within a few minutes (depending upon the component masses) of the GW detection. In the future, during the third observing run of LIGO and Virgo (O3), 
source classification information will be provided at multiple levels of refinement as parameter estimation results are made available.

\subsubsection{Skymaps and Luminosity Distances}
\label{localizationO2}

Currently, CBC sky localization probability maps (skymaps) for modeled searches are produced by two different algorithms, based on latency and sophistication: LALInference and BAYESTAR. LALInference uses stochastic sampling techniques for the entire parameter space of a CBC signal, such as Markov Chain Monte Carlo (MCMC) and nested sampling \citep{PhysRevD.91.042003}. Kernel density estimation is applied to the posterior samples to construct a smooth probability distribution from which the sky location and distance information is calculated. Although LALInference skymaps use the most complete description of the signal, sampling is computationally expensive, with a latency ranging from hours to days and weeks. BAYESTAR circumvents this issue by utilizing the fact that most of the information related to localization is captured by the arrival time, coalescence phase, and amplitude of the signal \citep{PhysRevD.93.024013}.\footnote{During the late stages of O2 with the Virgo detector, more complex matched-filter SNR time series from the detection pipelines were used in place of the point estimates.} As implemented in O2, the BAYESTAR likelihood is equivalent to that of LALInference. The marginalization is carried out via gaussian quadratures and lookup tables. Hence, the computation can be completed within a few seconds. Due to its highly parallel nature, a typical BAYESTAR skymap is computed in 30 s.

Both LALInference and BAYESTAR provide distance information in the skymaps. The distance is estimated from the moments of the posterior distance distribution conditioned on sky position.\footnote{See section 5 of \citep{0067-0049-226-1-10} for details on volume reconstruction.} In the case of LALInference the moments are calculated from a kernel density estimate trained on the posterior samples while for BAYESTAR the moments are calculated by numerical quadrature of the posterior probability distribution. 

As mentioned in section~\ref{unmodeledSearches}, burst triggers are generated by two algorithms, cWB and oLIB which produce their respective skymaps. The detection statistic of cWB is sensitive to the time delay in arrival of the signal at the detector sites and, thus, is a function of the sky position. The skymap is constructed based on the likelihood at each point in the sky (see \citealt{PhysRevD.93.042004} for details). The oLIB skymap algorithm, LALInferenceBurst, is similar to its CBC counterpart, LALInference, in the sense of being a template based search algorithm, except that it uses only sine-Gaussian templates. It reports a posterior in nine parameters where marginalization of parameters apart from sky position forms the skymap. Unlike CBC skymaps, burst skymaps do not contain distance information due to the lack of signal model.

Each skymap is distributed as a FITS file, a post-processed representation of the posterior samples to facilitate common queries and calculations regarding area and distance/volume when applicable. Consistent but non-numerically identical results are expected of integrals computed using FITS files versus MCMC with the posterior samples. Thus, luminosity distance medians in Table~\ref{tab:Distributedalerts} and in Table III of \citet{CBCcatalog} differ but agree to within about 5\% of the uncertainties in these quantities.

It is not unusual for both CBC and burst pipelines to identify the same astrophysical event, especially for heavier BBH mergers where the waveform is of shorter duration. In such cases, it is expected that the modeled CBC skymaps have smaller sky localizations. While there might be differences in the sky localizations, most of the probability is contained around the true location (see \citealt{doi:10.1093/mnrasl/slw239} for a comparative study).

\begin{figure}
\includegraphics[width=0.50\textwidth]{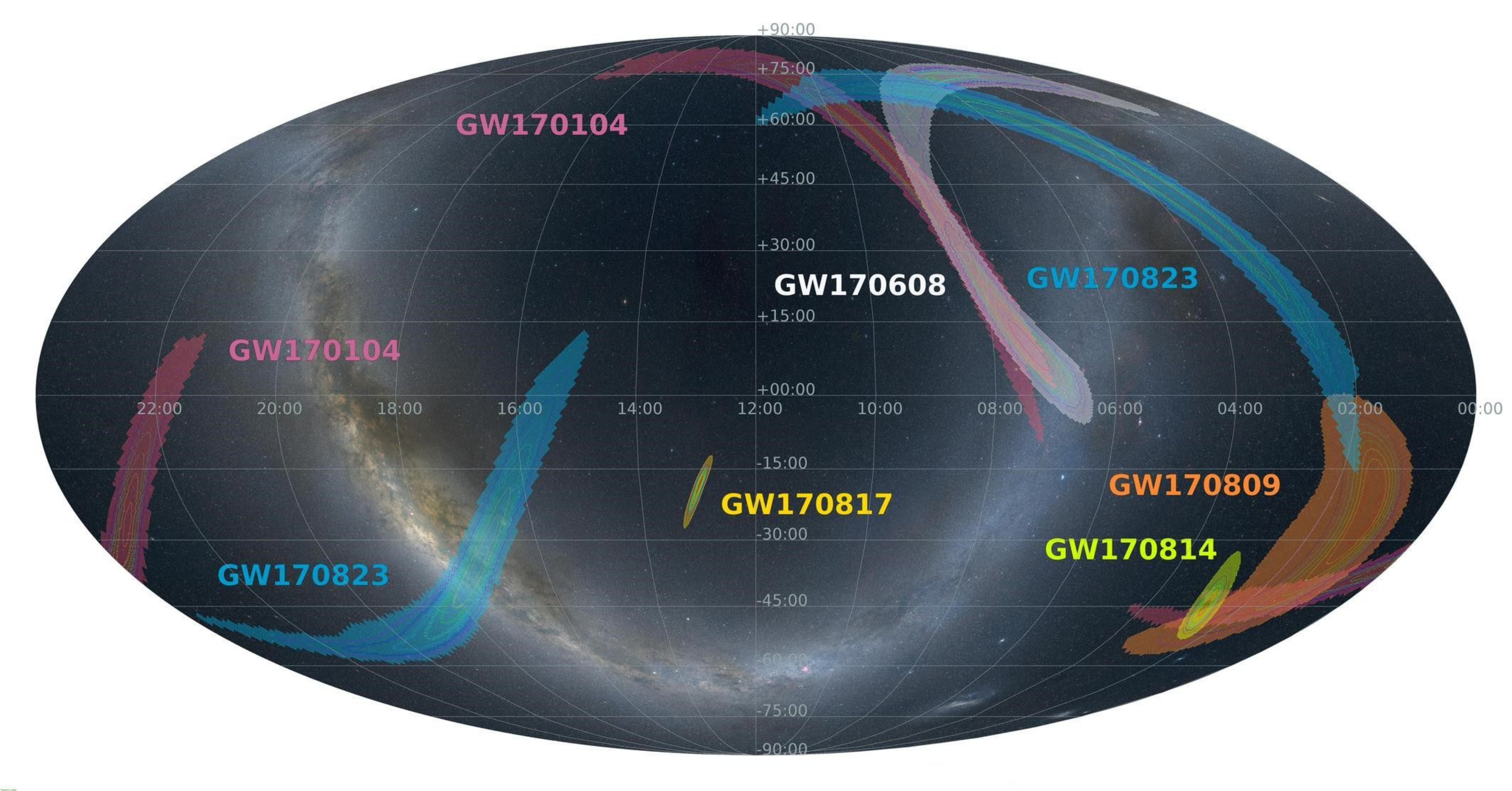}
\includegraphics[width=0.50\textwidth]{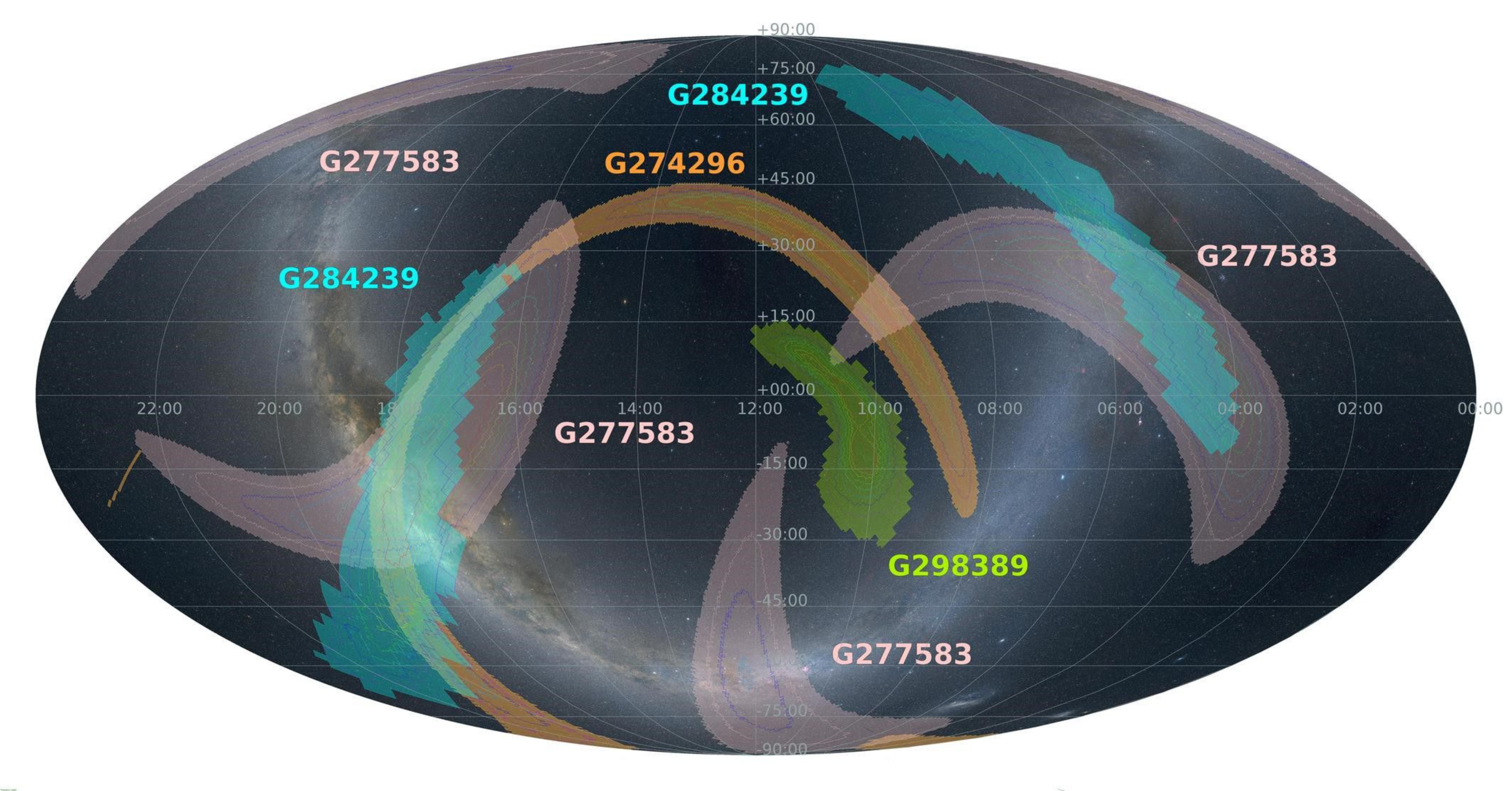}
\includegraphics[width=0.50\textwidth]{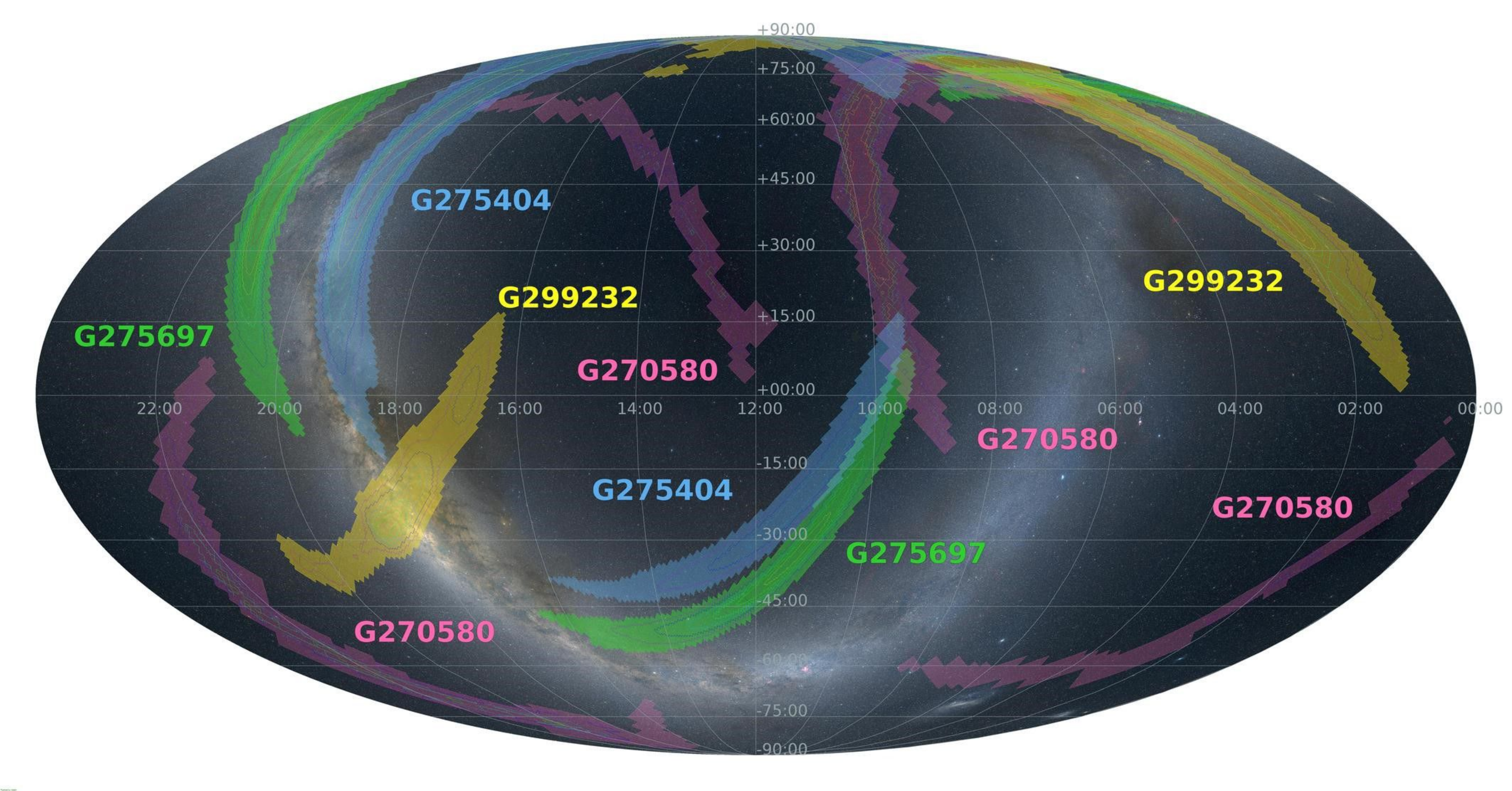}
\caption{Distributed low-latency O2 skymaps in ICRS coordinates - Mollweide projection. 
 The shaded areas correspond to the confidence region that encloses 90$\%$ of the localization probability.
The inner lines define the target regions at a 10$\%$ confidence level with changing color scheme at every $10\%$ increase in confidence.
The upper panel shows confident events, the middle panel GW candidates consistent with noise, 
and the bottom panel the triggers rejected by the offline analysis.
}
\label{fig:O2skymaps}
\end{figure}

\begin{figure}
\includegraphics[width=0.50\textwidth]{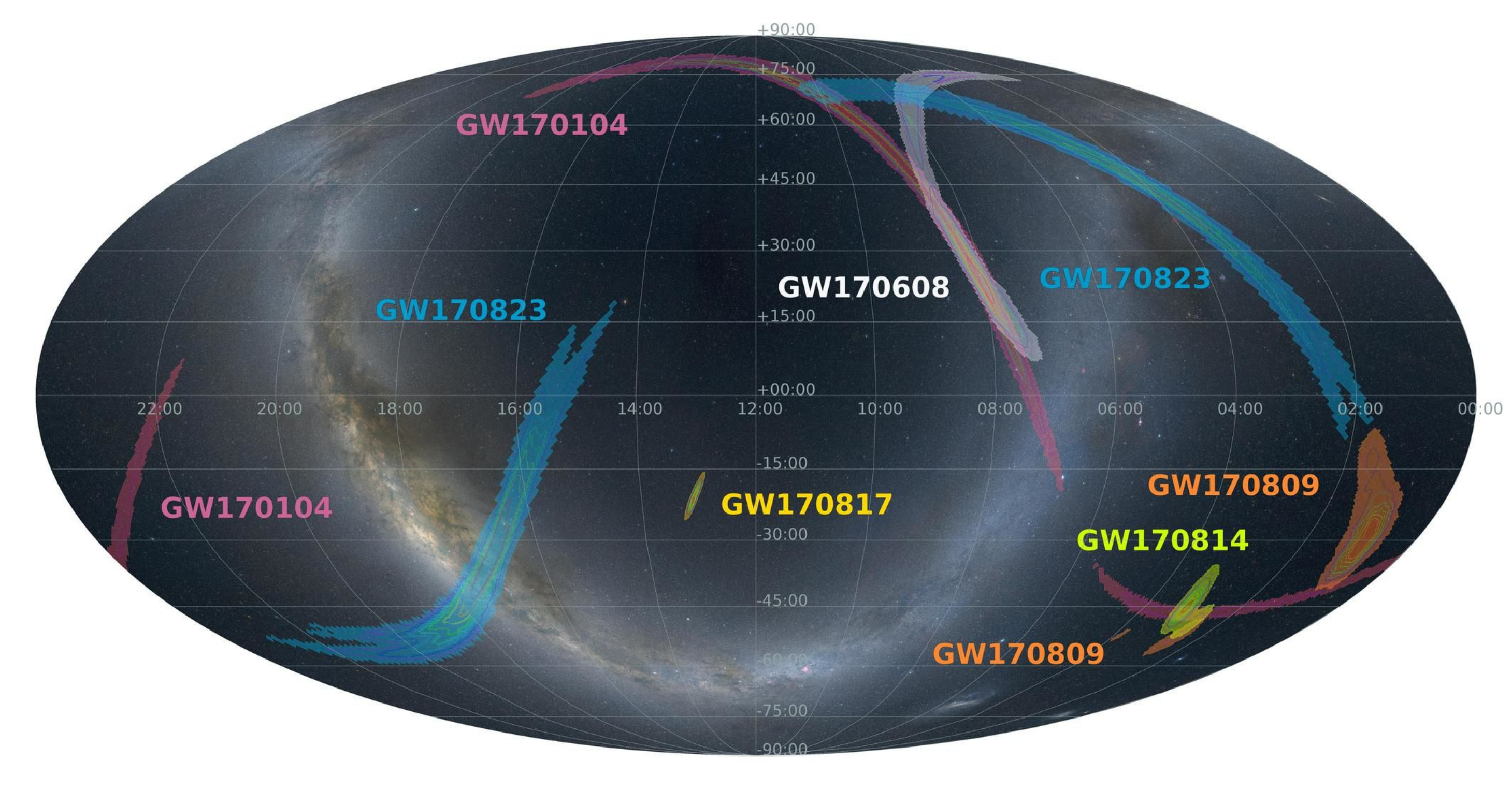}
\caption{Offline O2 sky localizations for the confident events published in the official LIGO and Virgo catalog - Mollweide projection. 
The shaded areas define the 90$\%$ confidence levels.
The inner lines define the target regions at a 10$\%$ confidence level with changing color scheme at every $10\%$ increase in confidence.
}
\label{fig:O2skymaps_final}
\end{figure}

Figure~\ref{fig:O2skymaps} shows the skymaps distributed during O2 using the low-latency algorithms discussed above. The upper panel shows the sky localizations of the high significance candidates (FAR $<$ 1/(100 yr) = $3 \times 10^{-10}$ Hz), which are confident GW events. Their corresponding refined sky localizations are shown in Figure~\ref{fig:O2skymaps_final}.  Taking into account the 90$\%$ confidence region, the initial sky localizations are largely consistent with the final localizations, with GW170814 being the exception (see section~\ref{antenna}).

A number of candidates (principally from burst pipelines) are consistent with noise and are not considered to be GW events. These are shown in the middle panel of Figure~\ref{fig:O2skymaps}. The bottom panel of the same figure shows the triggers rejected by offline analysis.

\subsubsection{Three-Detector Observations}
\label{antenna}

The Virgo contribution to O2 is noteworthy in significantly improving the localizations of the events GW170809, GW170814 and GW170817.\footnote{Virgo data also contributed to good localization of the event GW170818 that was found by the offline search \citep{CBCcatalog}.} As an example, for GW170814, the 50$\%$ sky localization area is confined in a single region of tens of square degrees in the southern hemisphere. We note that Virgo data was used to produce updated skymaps of GW170823 \citep{GCN21661,GCN21693} soon after identification of the signal. Subsequent data validation studies identified problems with the Virgo data around GW170823 which made it unreliable for use in parameter estimation: the final sky localization relies only on the LIGO data.

A note should be made regarding the initial and refined skymaps of GW170814 \citep{2017PhRvL.119n1101A,PhysRevLett.119.161101}: it is expected that the initial and updated skymaps for compact binary coalescence events are similar unless there are significant changes in data calibration, data quality or glitch treatment or low-latency parameter estimations.  

We observed a significant shift between the first GW170814 sky localization area and its update \citep{GCN21493, GCN21934}. At the time of GW170814, the Virgo power spectral density (PSD) changed significantly with respect to the estimated PSD used to precompute the template bank by the GstLAL online search. Consequently, the phase of the whitening filter for Virgo was no longer cancelled out by the templates. This resulted in a Virgo residual phase, which produced a phase shift in the Virgo SNR time series causing an east shift of the 50\% credible region of the sky localization area.

\begin{figure}
\includegraphics[width=0.50\textwidth]{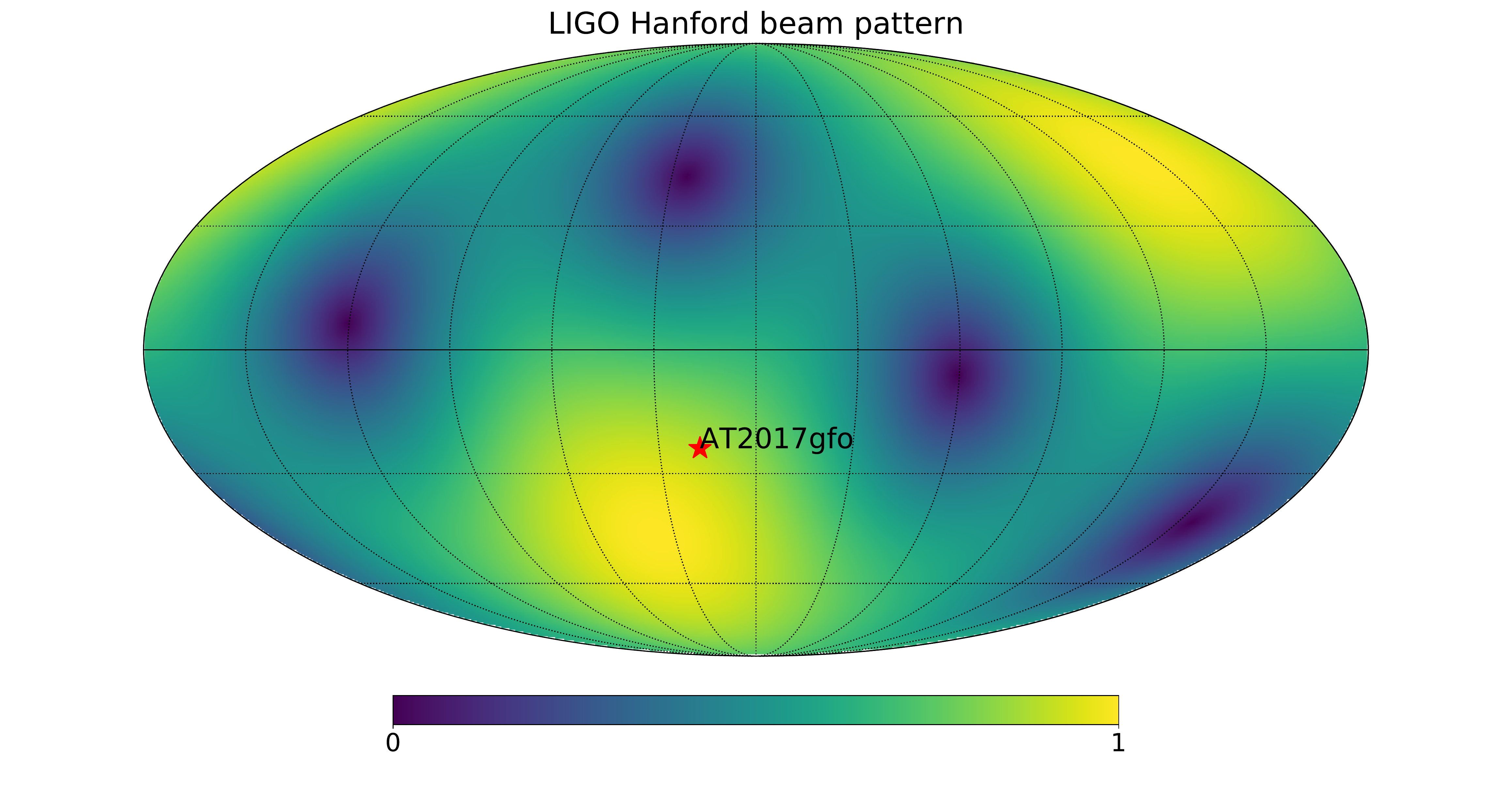}
\includegraphics[width=0.50\textwidth]{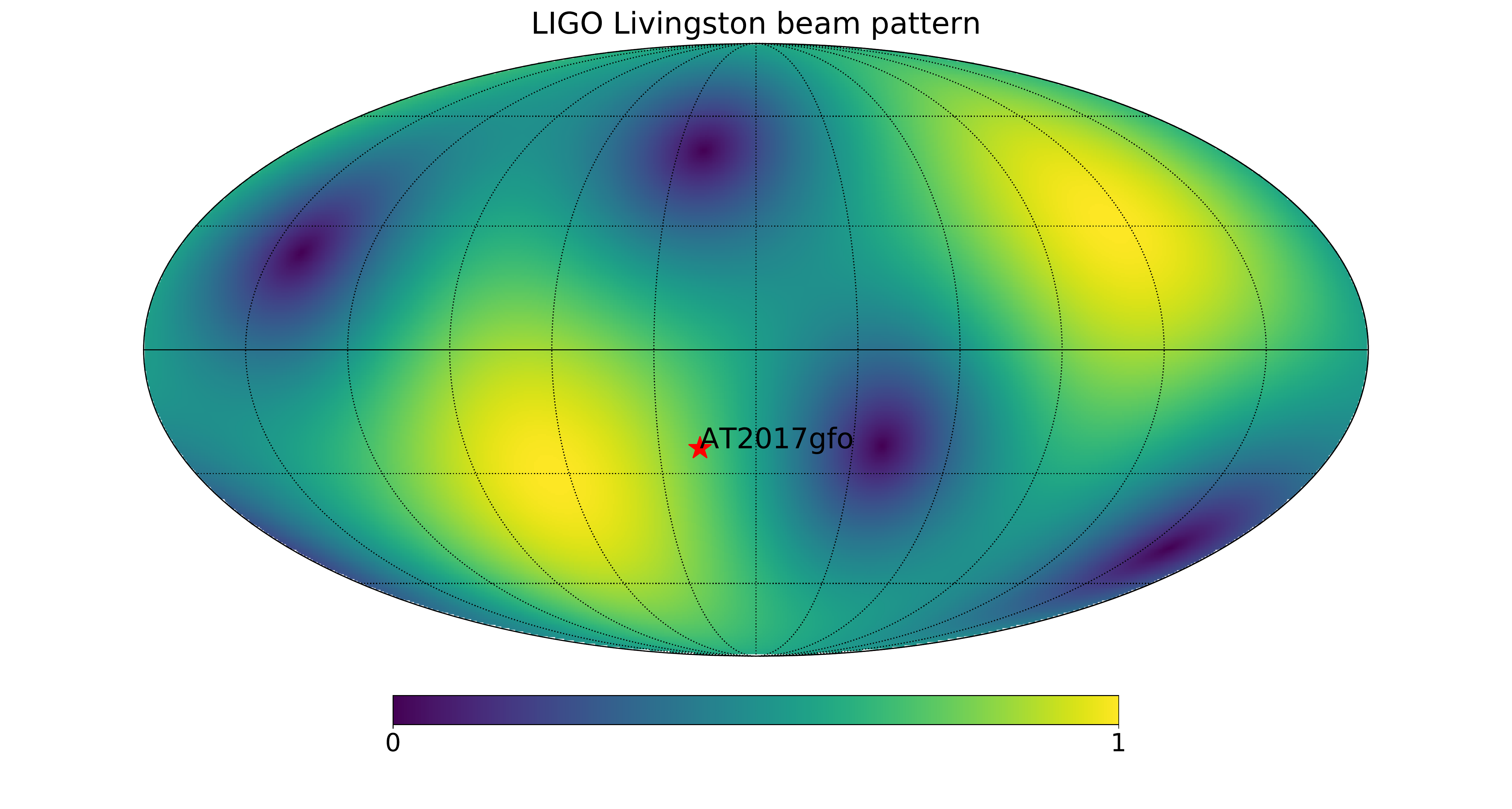}
\includegraphics[width=0.50\textwidth]{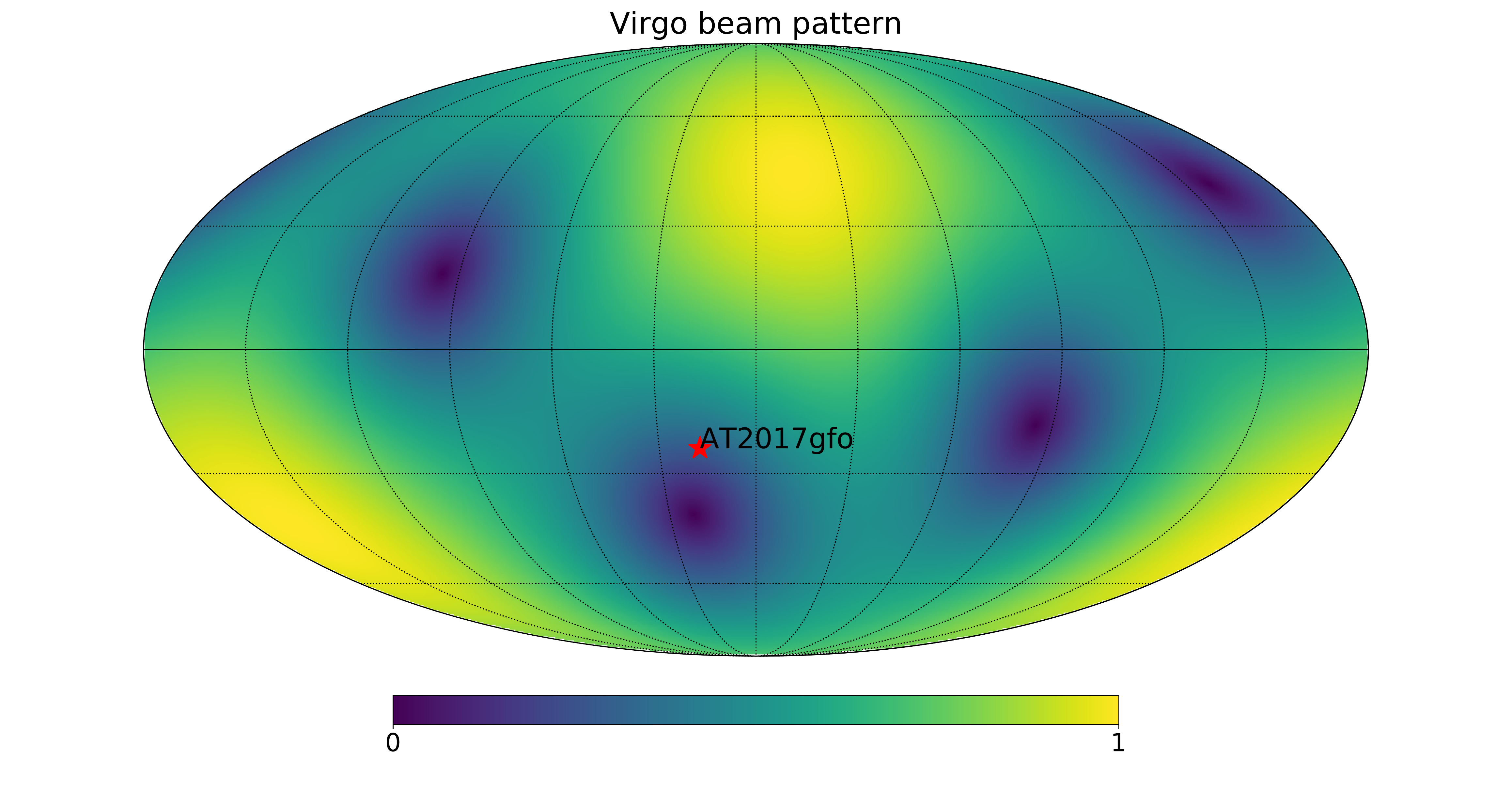}
\caption{Antenna patterns of the GW detectors in the network at the time of the GW170817 event
\textit{Upper}. LIGO-Hanford. \textit{Middle}. LIGO-Livingston. \textit{Bottom}. Virgo. ICRS coordinates - Mollweide projection. 
The position of the optical transient AT 2017gfo is indicated with a red star.
The color indicates the strength of the response with 
yellow being the strongest and blue being the weakest. At the location of GW170817, the antenna pattern amplitude for V is 2.5 to 3 times lower than for H and L.}
    
\label{fig:HLV_antenna}
\end{figure}

The antenna patterns for the three detectors are shown in Figure~\ref{fig:HLV_antenna}. The patterns represent the sensitivity of a detector to an event on the sky. The generic L-shaped detectors are most sensitive to signals coming from a direction perpendicular to the plane of its arms; this explains the two antipodal regions of maximum sensitivity. The plane of the arms of the detector form the least sensitive region. In this plane, the detectors are insensitive to the signal if directed along a diagonal of the L shape giving four islands of insensitivity.

\subsection{Electromagnetic/neutrino follow-up activity of gravitational waves alerts}
\label{MMfollow-up}

Searches for electromagnetic/neutrino counterparts employed a variety of observing strategies, including archival analysis, prompt searches with all-sky instruments, wide-field tiled searches, targeted searches of potential host galaxies, and deep follow-up of individual sources. They took into account the properties of the instruments, their observational capabilities (e.g., location on Earth for ground-based telescopes, pointing strategy for space-based instruments), and the characteristics of possible counterparts within their sensitivity band. For instance, the GW170817/GRB 170817A observational campaign perfectly illustrates the different observing strategies that led to the identification of the associated multi-wavelength electromagnetic emission: the independent identification of the sGRB by Fermi, the all-sky archival searches and wide-field of view follow-up, the discovery of the host galaxy and the optical GW counterpart by galaxy catalog targeted searches, the spectroscopic characterization of the optical counterpart, and the identification of X-ray and radio counterparts with deep follow-up \citep{2017ApJ...848L..12A}. Similar strategies were applied to the other triggers sent in O2, but on a more modest scale.

\paragraph{All-sky searches}\ Using temporal and spatial information, the large survey instruments (thousands of square degrees, up to over half of the sky) matched their independent transient database with the GW event or performed sub-threshold investigation near the trigger time and localization area of GW events. These strategies were used by neutrino detectors (ANTARES and IceCube, \citealt{GCN20866, GCN20861}), and high energy instruments like HAWC, \textit{Fermi}/GBM, CGBM/Calet, \textit{Swift}/BAT, \textit{AstroSAT}/CZTI, INTEGRAL \citealt{GCN21448, GCN20399} and \citealt{GCN21711, GCN21486, GCN21622, GCN20648, GCN270580}.  Wide-field optical survey data like SVOM/mini-GWAC \citep{GCN20745}, Pan-STARRS \citep{GCN20518}, iPTF \citep{GCN20398}, provided pre-merger images and for prompt/early emission. The time window around the GW candidate used to search for the EM counterpart is defined on the basis of the type of GW event and the counterpart properties expected in a specific band. For GRBs, it typically covered a few seconds to minutes. In the case of neutrinos, a window of  $\pm 500$ s around the merger  was used to search for neutrinos associated with prompt and extended gamma-ray emission \citep{2011APh....35....1B},  and a longer 14-day time window after the GW detection to cover predictions of longer-lived emission processes \citep{2017ApJ...849..153F}.

\smallbreak

\paragraph{Tiled and galaxy catalog targeted searches}\ The LIGO/Virgo alerts enabled EM follow-up campaigns by scanning large portions of the gravitational-wave sky localization error box or by targeting galaxies located within it \citep{0004-637X-820-2-136}. In the case of CBC triggers, the 3D sky-distance maps (see section~\ref{localizationO2}) were used to set observational strategies using the available galaxy 
catalogs \citep{2012A&A...539A.124L,2013ApJ...767..124N,0004-637X-784-1-8}. Other strategies were also employed, for example selecting strong-lensing galaxy clusters that lie within the 90\% credible region (GLGW Hunters \citealt{GCN21697}). This early follow-up of gravitational waves generally lasted tens of hours after the alert with observations from X-ray telescopes (such as \textit{Swift}/XRT \citealt{GCN21503} and MAXI/GSC \citealt{GCN21494}) as well as ground-based telescopes (e.g., MASTER \citealt{GCN21499}, PAN-STARRs \citealt{GCN21488}, DESGW/DECam \citealt{GCN21789}, GRAWITA/REM \citealt{GCN21495}, J-GEM/Subaru Hyper Suprime-Came \citealt{GCN21497}, GRAWITA/VST \citealt{GCN21498}, Las Cumbres/2-m \citealt{GCN21860}, KU/LSGT \citealt{GCN21885}, Pirate/0.43cm \citealt{GCN21888}). 

\smallbreak

\paragraph{Deep follow-up and classification of the counterpart candidates}\ After identification of potential counterparts, further classification was pursued with narrow field-of-view and sensitive instruments.
The large numbers of candidates and limited availability of larger instruments were two of the difficulties in counterpart 
identification. 

During the O2 follow-up campaign, most of the X-ray and optical candidates were classified through spectroscopic observations, which identified contaminants such as Galactic novae, supernovae, and active galactic nuclei (see e.g., observation campaign of GW170814 campaign by \citealt{GCN21512}).\footnote{https://gcn.gsfc.nasa.gov/other/G297595.gcn3}

During the GW170104 follow-up campaign, the ATLAS survey reported a rapidly fading optical source called ATLAS17aeu 
in coincidence with the GW skymaps, $\sim$21.5 hours after the GW trigger time.  Deeper investigations with a collective 
approach demonstrated that ATLAS17aeu was the afterglow of a long, soft gamma-ray burst GRB 170105A, 
unrelated with GW170104 \citep{2017ApJ...845..152B, 2017ApJ...850..149S, 2018arXiv180703681M}. 
This was an example where a coordinated follow-up of a GW event led to a serendipitous observation of an unrelated interesting event in time-domain astronomy.

\smallbreak

\paragraph{Long term follow-up}\ The long term follow-up of gravitational waves is also indispensable. 
One of the main challenges in radio follow-up of GW events is the association of the counterpart candidate found in the GW source localization region with the GW event, primarily due to a lack of temporal coincidence \citep{2015MNRAS.450.1430H, 2016ApJ...829L..28P}. 
However, the science return is potentially immense for such long term follow-up. For example, long-term X-rays, optical, and radio monitoring of GW170817 provided constraints on jet emission scenarios and models \citep{2016ApJ...831..190H, 2017ApJ...848L..25H, 2017arXiv171111573M, 2017arXiv171005435H, 2018arXiv180103531M,2018arXiv180800469G}.

\smallbreak

\begin{figure}
\includegraphics[width=0.5\textwidth]{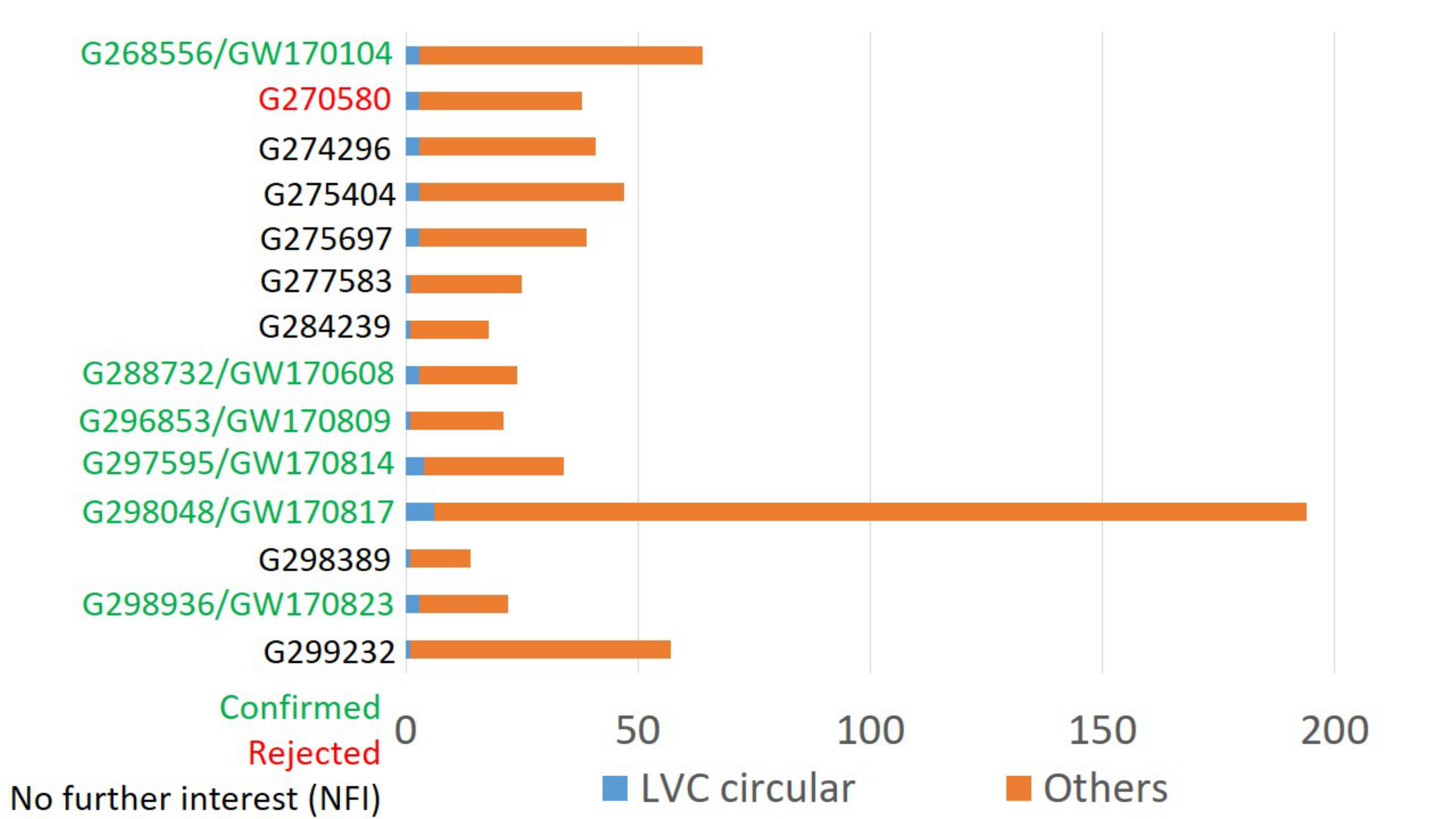}
 \caption{Summary of the exchange of information between \LVC{} and its partners showing the extensive follow-up activity. The color code for the alerts refers to the status of the alerts (confident, rejected, no further interest triggers).}
 \label{fig:GCNactivity}
\end{figure}

Figure~\ref{fig:GCNactivity} summarizes the exchange of information between \LVC{} and observing partners showing the extensive follow-up activity. More than 20 circulars were generated during each follow-up campaign. From Table~\ref{tab:Distributedalerts}, we note that the two oLIB events (G284239 and G298389) were the least followed due to their lower significance and higher latency in delivery of the initial skymaps. Also, more than 40 GCNs were generated for GW candidates that included a neutron star as one of the binary components (G275404, G275697, G298048, and G299232). G299232 was more extensively followed due to its classification as a potential NS-BH than confident detections like GW170814 or GW170608, which were BBH coalescences. This underscores the importance of source classification during O3, when observers might want to allocate their valuable resources in the most efficient manner possible.

From the EM follow-up activity side, no significant counterpart associated with BBH events was discovered; the most promising candidate was a weak gamma-ray transient found by AGILE during GW170104 lasting 32 ms and occuring 0.5 s before the GW event \citep{2017ApJ...847L..20V} but not confirmed by other instruments.

\smallbreak

\section{Conclusion}
\label{sec:conclu}
The O2 follow-up campaign of GW candidate events was a comprehensive effort of collaborating groups in astronomy and astro-particle physics. This effort was enabled by GW alerts distributed by the LIGO and Virgo collaborations. The triggers were produced by modeled searches for compact binary coalescences and unmodeled searches for transients such as the core-collapse of massive stars or neutron star instabilities.

During O2, 14 alerts were distributed with latencies ranging from 22 minutes up to 16 hours, most of them in less than an hour. Six events were declared as confident GW events associated to the merger of black holes or neutrons stars. The latency in sending alerts was dominated by human vetting of the candidate, which was necessary to validate data quality information beyond the ability of automated checks in place. O2 alerts, with false alarm rates less than one per two months, were distributed via the private GCN (Gamma-ray Coordinates Network) and contained GW information required for an efficient follow-up: the event time, sky localization probability map, and estimated false alarm rates. For compact binary merger candidates, skymaps with a third dimension (distance), the probability of the system to contain a neutron star and the probability to be electromagnetically bright were provided. The sky localization area of distributed events, which was hundreds to thousands of square degrees with the two LIGO interferometers, was dramatically reduced at the end of the campaign with the inclusion of Virgo.

The O2 follow-up program enabled the first combined observation of a neutron-star merger in gravitational waves (GW170817), gamma rays (GRB 170817A), and at optical wavelengths (AT 2017gfo). Together with the identification of the host galaxy and the subsequent observations of the X-ray and radio counterparts, the data collected on this event has yielded multiple groundbreaking insights into kilonova physics, the origin of heavy elements, the nature of neutron-star matter, cosmology, and basic physics.
The success of GW170817 and the larger O2 follow-up campaign demonstrates the importance of a coordinated multi-wavelength follow-up program for O3 and beyond.

Future priorities of the LIGO/Virgo Collaborations include further reduction in latency of GW alerts. The first hours after BNS mergers are crucial for observing the early X-rays, UV and optical emissions with space and ground instruments. For example, in the case of GW170817, the five hours delay in distributing skymaps, due to human intervention required to window out a glitch in LIGO-Livingston data, prevented the discovery of early emission, which could have revealed more about the the merger remnant and emission processes.

Beginning with the O3 observing run, LIGO and Virgo will issue public alerts\footnote{https://emfollow.docs.ligo.org/userguide/index.html}. There will be automated \textit{preliminary} notices generated based on the low-latency analysis. The FAR threshold for issuing \textit{preliminary} notices will be set so that it yields alerts with high confidence (at the level of 90\%) of having astrophysical origin for GW source types with populations that have been reliably measured until now. For other transient sources that have not been detected yet, the threshold will be lower, at the level of 1 per year. These will be followed by human vetted \textit{initial} notices or \textit{retraction} notices with a latency on the order of tens of minutes for high-interest candidates and within hours for more routine detections. We expect an increase in the number of GW events; BBH merger candidates will dominate by one order of magnitude from a few per week to a few per month whereas the BNS coalescence candidates are anticipated to occur a few times per year \citep{2010CQGra..27q3001A, 2016PhRvX...6d1015A, PhysRevLett.119.161101, 2018LRR....21....3A, CBCcatalog}. Both the increase in the number of significant candidate events and the need to reduce the latency of sending alerts will require an updated alert distribution infrastructure with a nearly fully-automated vetting protocol.

We have detailed the transient identification and alert systems utilized during the second LIGO/Virgo observing run. This work played a crucial role in ushering in the era of gravitational wave multi-messenger astronomy.

\acknowledgments
The authors gratefully acknowledge the support of the United States
National Science Foundation (NSF) for the construction and operation of the
LIGO Laboratory and Advanced LIGO as well as the Science and Technology Facilities Council (STFC) of the
United Kingdom, the Max-Planck-Society (MPS), and the State of
Niedersachsen/Germany for support of the construction of Advanced LIGO 
and construction and operation of the GEO600 detector. 
Additional support for Advanced LIGO was provided by the Australian Research Council.
The authors gratefully acknowledge the Italian Istituto Nazionale di Fisica Nucleare (INFN),  
the French Centre National de la Recherche Scientifique (CNRS) and
the Foundation for Fundamental Research on Matter supported by the Netherlands Organisation for Scientific Research, 
for the construction and operation of the Virgo detector
and the creation and support  of the EGO consortium. 
The authors also gratefully acknowledge research support from these agencies as well as by 
the Council of Scientific and Industrial Research of India, 
the Department of Science and Technology, India,
the Science \& Engineering Research Board (SERB), India,
the Ministry of Human Resource Development, India,
the Spanish  Agencia Estatal de Investigaci\'on,
the Vicepresid\`encia i Conselleria d'Innovaci\'o, Recerca i Turisme and the Conselleria d'Educaci\'o i Universitat del Govern de les Illes Balears,
the Conselleria d'Educaci\'o, Investigaci\'o, Cultura i Esport de la Generalitat Valenciana,
the National Science Centre of Poland,
the Swiss National Science Foundation (SNSF),
the Russian Foundation for Basic Research, 
the Russian Science Foundation,
the European Commission,
the European Regional Development Funds (ERDF),
the Royal Society, 
the Scottish Funding Council, 
the Scottish Universities Physics Alliance, 
the Hungarian Scientific Research Fund (OTKA),
the Lyon Institute of Origins (LIO),
the Paris \^{I}le-de-France Region, 
the National Research, Development and Innovation Office Hungary (NKFIH), 
the National Research Foundation of Korea,
Industry Canada and the Province of Ontario through the Ministry of Economic Development and Innovation, 
the Natural Science and Engineering Research Council Canada,
the Canadian Institute for Advanced Research,
the Brazilian Ministry of Science, Technology, Innovations, and Communications,
the International Center for Theoretical Physics South American Institute for Fundamental Research (ICTP-SAIFR), 
the Research Grants Council of Hong Kong,
the National Natural Science Foundation of China (NSFC),
the Leverhulme Trust, 
the Research Corporation, 
the Ministry of Science and Technology (MOST), Taiwan
and
the Kavli Foundation.
The authors gratefully acknowledge the support of the NSF, STFC, MPS, INFN, CNRS and the
State of Niedersachsen/Germany for provision of computational resources.

This article has been assigned the document number LIGO-P1800255.

\end{document}